\documentclass[amsmath,amssymb,widetext,10pt,a4paper]{revtex4}
\usepackage{graphicx}
\usepackage{bm}
\def\beq{\begin{equation}}
\def\eeq{\end{equation}}

\begin{document}

\title{Current-driven switching of magnetisation- theory and experiment}

\author{D. M.  Edwards}
\affiliation{Department of Mathematics, Imperial College, London SW7
  2BZ, U.K.}
\author{ F. Federici}
\affiliation{NEST-INFM and Classe di Scienze, Scuola Normale
  Superiore, Piazza dei Cavalieri 7, 56126 Pisa, Italy}

\maketitle

\section{Introduction}\label{intro}

Recently there has been a lot of interest in magnetic nanopillars of
10-100 nm in diameter. The pillar is a metallic layered structure
with two ferromagnetic layers, usually of cobalt, separated by a
non-magnetic spacer layer, normally of copper. Non-magnetic leads
are attached to the magnetic layers so that an electric current may
be passed through the structure. In the simplest case the pillar may
exist in two states, with the magnetisation of the two magnetic
layers parallel or anti-parallel. The state of a pillar can be read
by measuring its resistance, this being smaller in the parallel
state than in the anti-parallel one. This dependence of the
resistance on magnetic configuration is the giant magnetoresistance
(GMR) effect \cite{1}. A dense array of these nanopillars could form
a magnetic memory for a computer. Normally one of the magnetic
layers in a pillar is relatively thick and its magnetisation
direction is fixed. In order to write into the memory the
magnetisation direction of the second thinner layer must be
switched. This might be achieved by a local magnetic field of
suitable strength and methods have been proposed \cite{2} for
providing such a local field by currents in a criss-cross array of
conducting strips. However an alternative, and potentially more
efficient method, proposed by Slonczewski \cite{3} makes use of a
current passing up the pillar itself. Slonczewski's effect relies on
``spin transfer'' and not on the magnetic field produced by the
current which in the nanopillar geometry is ineffective. The idea of
spin-transfer is as follows. In a ferromagnet there are more
electrons of one spin orientation than of the other so that current
passing through the thick magnetic layer (the polarising magnet)
becomes spin-polarised. In general its state of spin-polarisation
changes as it passes through the second (switching) magnet so that
spin angular momentum is transferred to the switching magnet. This
transfer of spin angular momentum is called spin-transfer torque
and, if the current exceeds a critical value, it may be sufficient
to switch the direction of magnetisation of the switching magnet.
This is called current-induced switching.

In the next section we show how to calculate the spin-transfer torque
for a simple model.

\section{Spin-transfer torque in a simple model}\label{due}

For simplicity we consider a structure of the type shown in Fig.
\ref{fig1}, where {\bf p} and {\bf m} are unit vectors in the
direction of the magnetisations. This models the layered structure
of the pillars used in experiments but the atomic planes shown are
considered to be unbounded instead of having the finite
cross-section of the pillar. This means that there is translational
symmetry in the $x$ and $z$ directions. The structure consists of a
thick (semi-infinite) left magnetic layer (polarising magnet), a
non-magnetic metallic spacer layer, a thin second magnet (switching
magnet) and a semi-infinite non-magnetic lead. In the simplest model
we assume the atoms form a simple cubic lattice, with lattice
constant $a$, and we adopt a one-band tight-binding model with
hopping Hamiltonian

\begin{equation}
H_0=t\sum_{{\bf k}_{\parallel}\sigma}
\sum_{n}c^{\dagger}_{k_{\parallel} n\sigma} c_{k_{\parallel}
n-1\sigma}\,\, +\,\, {\rm h. c.}. \label{hamiltonian}
\end{equation}

Here $c^{\dagger}_{k_{\parallel} n\sigma}$ creates an electron on
plane $n$ with two-dimensional wave-vector ${\bf k}_{\parallel}$ and
spin $\sigma$, and $t$ is the nearest-neighbour hopping integral.

In the tight-binding  description the operator for spin angular
momentum current between planes $n-1$ and $n$, which we require to
calculate spin-transfer torque, is given by
\begin{equation}
{\bf j}_{n-1}=-\frac{{\rm i}t}2\sum_{{\bf
    k}_{\parallel}}\left(c^{\dagger}_{k_{\parallel},n,\sigma\uparrow},
c^{\dagger}_{k_{\parallel},n,\downarrow}\right){\bm\sigma}
\left(c_{k_{\parallel} ,n-1,\uparrow}, c_{k_{\parallel}
,n-1,\downarrow} \right)^{\dagger} +\,\, {\rm h. c.}. \label{gamma}
\end{equation}
Here ${\bm\sigma}=\left(\sigma_x,\sigma_y,\sigma_z\right)$ where
the components are Pauli matrices. Eq. (\ref{gamma}) yields the
charge current $j_{n-1}^{\rm c}$ if $\frac12{\bm\sigma}$ is
replaced by a unit matrix multiplied by the number $e/\hbar$, where
$e$ is the electronic charge (negative). All currents flow in the
$y$ direction, perpendicular to the layers, and the components of
the vector ${\bf  j}$
 correspond to transport of $x$, $y$ and $z$ components of spin. The
 justification of Eq. (\ref{gamma}) for ${\bf j}_{n-1}$ relies
 on an equation of continuity, as pointed out in Section \ref{quattro}.

To define the present model completely we must supplement the
hopping Hamiltonian $H_0$ by specifying the on-site potentials in
the various layers. For simplicity we assume the on-site potential
for both spins in non-magnetic layers, and for majority spin in
ferromagnetic layers, is zero. We assume an infinite exchange
splitting in the ferromagnets so that the minority spin potential in
these layers is infinite. Thus minority spin electrons are
completely excluded from the ferromagnets. Clearly the definition of
majority and minority spin relate to spin quantisation in the
direction of the local magnetisation. We take $\alpha=0$, so that
the magnetisation of the switching magnet is in the z direction and
 take $\theta=\psi$, where $\psi$ is the angle
between the magnetisations.

To describe spin transport in the structure we adopt the generalised
Landauer approach of Waintal {\it et. al.} \cite{4}. Thus the
structure is placed between two reservoirs, one on the left and one
on the right, with electron distributions characterised by Fermi
functions $f(\omega-\mu_{\rm L})$, $f(\omega-\mu_{\rm R})$
respectively. The system is then subject to a bias voltage $V_{\rm
b}$ given by $eV_{\rm b}=\mu_{\rm L}-\mu_{\rm R}$, the difference
between the chemical potentials. We discuss the ballistic limit
where scattering occurs only at interfaces, the effect of impurities
being negligible. We label the atomic planes so that $n=0$
corresponds to the last atomic plane of the polarising magnet. the
planes of the spacer layer correspond to $n=1,2\, \cdot\cdot\,N $
and $n=N+1$ is the first plane of the switching magnet.

Consider first an electron incident from the left with wave-function
$|k,maj\rangle$, where $k>0$, which corresponds to a Bloch wave
$|k\rangle = \sum_{n}{\rm e}^{{\rm i}kna}|{\bf
k}_{\parallel}n\rangle$ with majority spin in the polarising magnet.
In this notation the label ${\bf k}_{\parallel}$ is suppressed. The
particle is partially reflected by the structure and finally emerges
as a partially transmitted wave in the lead, with spin $\uparrow$
corresponding to majority spin in the switching magnet. Thus the
wave-function is of the form
\begin{equation}
|P_k\rangle =|k,maj\rangle + B|-k,maj\rangle
\label{p}
\end{equation}
in the polariser and
\begin{equation}
|L_k\rangle =F|k,\uparrow\rangle
\label{p2}
\end{equation}
in the lead. A majority spin in either ferromagnet  enters or leaves
the spacer without scattering, since in our simple model there is no
potential step. Also the minority spin wave-function entering a
ferromagnet is zero. The spacer wave-function may therefore be written
in two ways:
 \begin{equation}
|S_k\rangle =F|k,\uparrow\rangle+E \left({\rm e}^{-{\rm
i}k(N+1)a}|k,\downarrow\rangle- {\rm e}^{{\rm
i}k(N+1)a}|-k,\downarrow\rangle \right) 
\label{pp}
\end{equation}
or
 \begin{eqnarray}\label{ppp}
|S_k\rangle &=& |k,maj\rangle+B|-k,maj\rangle+D\left(|k,min\rangle-
|-k,min\rangle\right)\nonumber\\
&=& \cos\left(\psi/2\right)|k,\uparrow\rangle+\sin\left(\psi/2\right)
|k,\downarrow\rangle
+B\left[\cos\left(\psi/2\right)|-k,\uparrow\rangle+\sin\left(\psi/2\right)|-k,\downarrow\rangle\right]\\
&&+D\left[-\sin\left(\psi/2\right)|k,\uparrow\rangle+\cos\left(\psi/2\right)
|k,\downarrow\rangle+\sin\left(\psi/2\right)
|-k,\uparrow\rangle-\cos\left(\psi/2\right)|-k,\downarrow\rangle\right].\nonumber
\end{eqnarray}
On equating coefficients of $|k,\uparrow\rangle$,
$|k,\downarrow\rangle$, $|-k,\uparrow\rangle$,
$|-k,\downarrow\rangle$ in expressions (\ref{pp}) and (\ref{ppp})
we have four equations which may be solved for $B$, $D$, $E$, $F$.
In particular the transmission coefficient $T$ is given by
\begin{equation}
\label{T}
T=\left|F\right|^2=\frac{4\cos^2(\psi/2)\sin^2
  k(N+1)a}{\sin^4(\psi/2)+
4\cos^2(\psi/2)\sin^2k(N+1)a}.
\end{equation}
Similarly an electron incident from the right with wave-function
$|-k\uparrow\rangle$ in the lead is partially reflected and finally
emerges as a partially transmitted wave $F^{\prime}|-k,maj\rangle$ in
the polarising magnet. It is found that $F^{\prime}=F$ so that the
transmission coefficient is the same for particles from left or right.

The spin angular momentum current in a particular layer, which we
shall denote by S although it need not be the spacer layer, is the
sum of currents carried by left and right moving electrons. Thus we
have a Landauer-type formula \cite{5}
\begin{equation}
\label{landau}
{\bf j}_{\rm s}=\frac{a}{2\pi}\sum_{{\bf k }_{\parallel}}
\left\{\int_{k>0}{\rm d}k\left[\langle S_k|{\bf
  j}_{n-1}|S_k\rangle f(\omega-\mu_L)+\langle S_{-k}|{\bf
  j}_{n-1}|S_{-k}\rangle f(\omega-\mu_{R})\right]\right\}
\end{equation}
where $|S_k\rangle$, $|S_{-k}\rangle$ are wave-functions in the layer
considered corresponding to electrons incident from left and right,
respectively. Here $\omega$, the energy of the Bloch wave $k$, is
given by the tight-binding formula
\begin{equation}
\label{omega}
\omega=u_{{\bf k}_{\parallel}}+2t\cos ka
\end{equation}
where $u_{{\bf k}_{\parallel}}=2t(\cos k_xa+\cos k_za)$. We take
$t<0$ so that positive $k$ corresponds to positive velocity
$\hbar^{-1}\partial\omega/\partial k$ as we have assumed. The
current ${\bf j}_{\rm s}$ in layer $S$ calculated by Eq.
(\ref{landau}) does not depend on the particular planes $n-1$, $n$
between which it is calculated. On changing the integration variable
in Eq. (\ref{landau}) we find
\begin{equation}
\label{landau2} {\bf j}_{\rm s}=\frac{1}{2\pi} \sum_{{\bf
k}_{\parallel}}\int{\rm d}\omega\left[{\bf J}_{+}f(\omega-\mu_{\rm
  L})+{\bf J}_{-}f(\omega-\mu_{\rm R})
\right]
\end{equation}
where
\begin{equation}
\label{landau3}
{\bf J}_{\pm}=\frac{\langle S_{\pm k}|{\bf j}_{n-1}|S_{\pm
    k}\rangle}{-2t\sin ka}.
\end{equation}
Here $k=k(\omega,{\bf k}_{\parallel})$ is the positive root of Eq.
(\ref{omega}). Eq. (\ref{landau2}) may be written as
\begin{equation}
\label{landau33}
{\bf j}_{\rm s}=\frac{1}{4\pi}\sum_{{\bf k }_{\parallel}}
\int{\rm d}\omega\left\{
\left({\bf J}_{+}+{\bf J}_{-}\right)
\left[f(\omega-\mu_{\rm L})+f(\omega-\mu_{\rm R})\right]+
\left({\bf J}_{+}-{\bf J}_{-}\right)
\left[f(\omega-\mu_{\rm L})-f(\omega-\mu_{\rm  R})\right]
\right\}.
\end{equation}

Before discussing this spin current we briefly consider the charge
current $j^{\rm c}$, and we denote the analogues of ${\bf J}_{\pm}$ by
$J_{\pm}^{\rm c}$. Since the charge current is conserved
throughout the structure $J_{+}^{\rm c}$ and  $J_{-}^{\rm c}$ can be
calculated in different ways, {\it e.g.} in the lead for $J_{+}^{\rm c}$ and in the polariser for $J_{-}^{\rm c}$. Since
$T=\left|F\right|^2=\left|F^{\prime}\right|^2$ we find $J_{+}^{\rm
  c}+J_{-}^{\rm c}=0$ and for small bias $eV_{b}=\mu_{\rm L}-\mu_{\rm R}$
the charge current is given by
\begin{equation}
\label{current}
j^{\rm c}=\frac{2e^2V_{\rm b}}{h}\sum_{{\bf
k}_{\parallel}}T
\end{equation}
where the transmission coefficient $T$ is given by Eq. (\ref{T})
with $k=k(\mu,{\bf k}_{\parallel})$, $\mu$ being the common chemical
potential as $V_{\rm B}\rightarrow 0$. This is the well-known
Landauer formula \cite{5}.

The spin transfer torque on the switching magnet is given by
\begin{equation}
\label{torque}
{\bf T}^{\rm s-t}=\langle{\bf j}_{\rm spacer}\rangle-
\langle{\bf j}_{\rm lead}\rangle,
\end{equation}
where $\langle{\bf j}_{\rm spacer}\rangle$ and $\langle{\bf j}_{\rm
lead}\rangle$ are spin currents in the spacer and lead respectively.
For zero bias ($\mu_{\rm L}=\mu_{\rm
  R}$) there is clearly no charge current in the structure and
straight-forward calculation shows that all components of spin
current in the spacer and the lead vanish, except for a non-zero
$y$-spin current in the spacer. There is therefore a non-zero $y$
component of spin-transfer torque acting on the switching magnet for
zero bias, and its dependence on the angle $\psi$ between the
magnetisations is found to be approximately $\sin\psi$. This torque
is due to exchange coupling, analogous to an RKKY coupling, between
the two magnetic layers. This coupling oscillates as a function of
spacer thickness and tends to zero as the thickness tends to
infinity. For finite bias $V_{\rm B}$ the second term in the
integrand of Eq. (\ref{landau33}) comes into play. In general this
leads to finite $x$ and $y$ components of ${\bf T}^{\rm s-t}$
proportional to $V_{\rm b}$ (for small $V_{\rm b}$) whereas
$T_{z}^{\rm s-t}=0$. However for the special model considered here
with infinite exchange splitting in both ferromagnets it turns out
that $T_y^{\rm s-t}=0$. For this model the only non-zero component
of ${\bf T}^{\rm s-t}$ proportional to
 $V_{\rm b}$ is found to be
 \begin{equation}
\label{torquex}
T_x^{\rm s-t}=\frac{\hbar j^{\rm c}}{2|e|}\tan\frac{\psi}{2}
\end{equation}
where $j^{\rm c}$ is the charge current given by Eq. (\ref{current}).

Slonczewski \cite{3} originally obtained this result for the
analogous parabolic band model. From Eqs. (\ref{torquex}),
(\ref{current}) and (\ref{T}) it follows that $T_{x}^{\rm s-t}$
contains an important factor $\sin\psi$ although this does not
represent the whole angle dependence. Clearly, from Eq.
(\ref{torquex}), the torque proportional to bias remains finite for
arbitrarily large spacer thickness, in the ballistic limit. For this
model, with infinite exchange splitting, the torque is independent
of the thickness of the switching magnet.

From the results of this simple model we can infer a general form of
the spin-transfer torque ${\bf T}^{\rm s-t}$ which is independent of
the choice of coordinate axes. Thus we write
 \begin{equation}
\label{torqueperp}
{\bf T}^{\rm s-t}={\bf T}_{\perp}+{\bf T}_{\parallel}
\end{equation}
where
 \begin{eqnarray}
\label{torque2} {\bf T}_{\perp}&=&\left(g^{\rm ex}+g_{\perp} e
V_{\rm b}\right)({\bf
  m}\times{\bf p}) \nonumber\\
{\bf T}_{\parallel}&=&g_{\parallel} e V_{\rm b}{\bf m}\times({\bf
  p}\times{\bf m}).
\end{eqnarray}
With the choice of axes in Fig. \ref{fig1} ${\bf T}_{\parallel}$
corresponds to the $x$ component of torque, that is the component
parallel to the plane containing the magnetisation directions ${\bf
m}$ and ${\bf p}$. Similarly ${\bf T}_{\perp}$ corresponds to the
$y$ component of torque, this being perpendicular to the plane of
${\bf m}$ and ${\bf p}$. The modulus of both the vectors ${\bf
m}\times{\bf p}$ and ${\bf m}\times({\bf p}\times{\bf m})$ is
$\sin\psi$, so that the factors $g^{\rm ex}$, $g_{\perp}$ and
$g_{\parallel}$ are functions of $\psi$ which contain deviations
from the simple $\sin\psi$ behaviour. The bias-independent term
$g^{\rm ex}$ corresponds to the interlayer exchange coupling, as
discussed above, and henceforth we assume that the spacer is thick
enough for this term to be negligible. Sometimes the $\sin\psi$
factor accounts for most of the angular dependence of $T_{\perp}$
and $T_{\parallel}$ so that $g_{\perp}$ and $g_{\parallel}$ may be
regarded as constant parameters for the given structure. In the next
section we use Eqs. (\ref{torque2}) for the spin-transfer torque in
a phenomenological theory of current-induced switching of
magnetisation. This phenomenological treatment enables us to
understand most of the available experimental data. It is more usual
in experimental works to relate spin-transfer torque to current
rather than bias. However in theoretical work, based on the Landauer
or Keldysh approach, bias is more natural. In practice the
resistance of the system considered is rather constant (the GMR
ratio is only a few percent) so that bias and current are in a
constant ratio.

\section{Phenomenological treatment of current-induced switching of
  magnetisation}\label{tre}
\begin{figure}
\includegraphics[width=11cm]{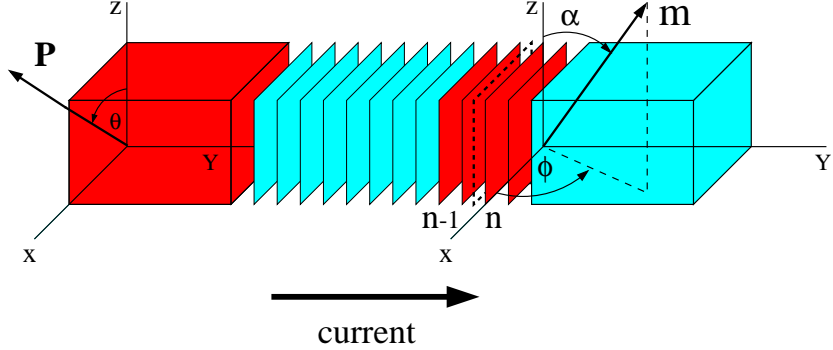}
\caption{Schematic picture of a magnetic layer structure for
  current-induced switching (magnetic layers are darker, non-magnetic
  layers lighter).}
\label{fig1}
\end{figure}

\begin{figure}
\includegraphics[width=8cm,angle=-90]{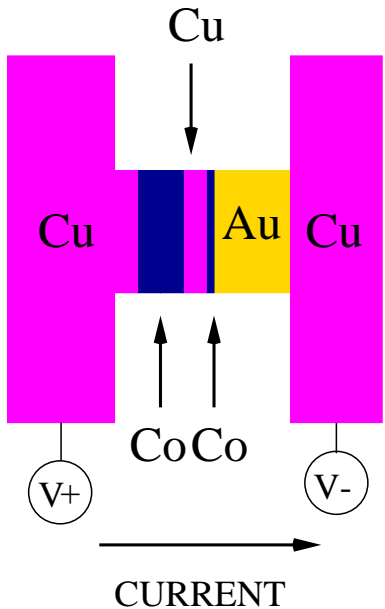}
\caption{Schematic picture of a junction in which current-induced
  switching is studied experimentally.}
\label{fig2}
\end{figure}

In this section we explore the consequences of the spin-transfer
torque acting on a switching magnet using a phenomenological Landau
Lifshitz equation with Gilbert damping (LLG equation). This is
essentially a generalisation of the approach used originally by
Slonczewski \cite{3} and Sun \cite{sun}. We assume that there is a
polarising magnet whose magnetisation is pinned in the $xz$-plane in
the direction of a unit vector ${\bf p}$, which is at general fixed
angle $\theta$ to the $z$-axis as shown in Fig. \ref{fig1}.
The pinning of the magnetisation of the polarising magnet can be due
to its large coercivity (thick magnet) or a strong uniaxial
anisotropy. The role of the polarising magnet is to produce a stream
of spin-polarised electrons, {\it i.e.} spin current, that is going
to exert a torque on the magnetisation of the switching magnet whose
magnetisation lies in the general direction of a unit vector ${\bf
m}$. The orientation of the vector ${\bf m}$ is defined by the polar
angles $\alpha$, $\phi$ shown in Fig. \ref{fig1}.
 There is a non-magnetic metallic layer inserted
between the two magnets
whose role is merely to separate magnetically the two magnetic  layers and
allow a strong charge current to pass. The total thickness of the
whole trilayer sandwiched between two non-magnetic leads must be
smaller than the spin diffusion length $l_{\rm sf}$ so that there
are no spin flips due to impurities or spin-orbit coupling. A
typical junction in which current-induced switching is studied
experimentally \cite{albert} is shown schematically in Fig.
\ref{fig2}. The thickness of the polarising magnet is 40nm, that of
the switching magnet 2.5nm and the non-magnetic spacer is 6nm thick.
The materials for the two magnets and the spacer are cobalt and
copper, respectively, which are those most commonly used. The
junction cross section is oval-shaped with dimensions
60nm$\times$130nm. A small diameter is necessary so that the torque
due to the Oersted field generated by a charge current of
$10^{7}$-$10^8$ A/cm$^2$, required for current-induced switching, is
much smaller than the spin-transfer torque we are interested in.

\begin{figure}
\includegraphics[width=8cm]{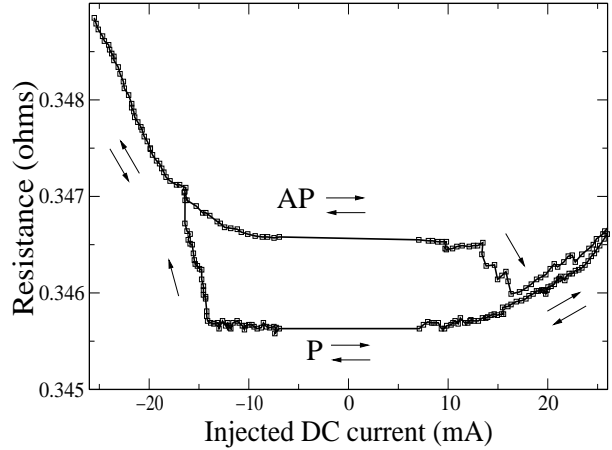}
\caption{Resistance vs current hysteresis loop (after Grollier {\it
et al.} \cite{16}). } \label{fig3}
\end{figure}

The aim of most experiments is to determine the orientation of the
switching magnet moment as a function of the current (applied bias)
in the junction. Sudden jumps of the magnetisation direction, {\it
i.e.} current-induced switching, are of particular interest. The
orientation of the switching magnet moment ${\bf m}$ relative to
that of the polarising magnet ${\bf p}$, which is fixed, is
determined by measuring the resistance of the junction. Because of
the GMR effect, the resistance of the junction is higher when the
magnetisations of the two magnets are anti-parallel than when they
are parallel, In other words, what is observed are hysteresis loops
of resistance versus current. A typical experimental hysteresis loop
of this type \cite{16} is
reproduced in Fig. \ref{fig3}.
It can be seen from Fig. \ref{fig3} that, for any given current, the
switching magnet moment is stationary (the junction resistance has a
well defined value), {\it i.e.} the system is in a steady state.
This holds everywhere on the hysteresis loop except for the two
discontinuities where current-induced switching occurs. As indicated
by the arrows jumps from the parallel (P) to anti-parallel (AP)
configurations of the magnetisation, and from AP to P
configurations, occur at different currents. It follows that in
order to interpret experiments which exhibit such hysteresis
behaviour, the first task of the theory is to determine from the LLG
equation all the possible states and then investigate their
dynamical stability. At the point of instability the system seeks
out a new steady state, {\it i.e.} a discontinuous transition to a
new steady state with the switched magnetisation occurs. We have tacitly
assumed that there is always a steady state available for the system
to jump to. There is now experimental evidence that this is not
always the case. In the absence of any stable steady state the
switching magnet moment remains permanently in the time-dependent
state. This interesting case is implicit in the phenomenological LLG
treatment and we shall discuss it in detail later.

In describing the switching magnet by a unique unit vector ${\bf m}$, we
assume that it remains uniformly magnetised during the switching
process. This is only strictly true when the exchange stiffness of the
switching magnet is infinitely large. It is generally a good
approximation as long as the switching magnet is small enough to
remain single domain, so that the switching occurs purely by rotation
of the magnetisation as in the Stoner-Wohlfarth theory \cite{17} of
field switching. This seems to be the case in many experiments
\cite{albert,16,18,19}

Before we can apply the LLG equation to study the time evolution of
the unit vector ${\bf m}$ in the direction of the magnetisation of
the switching magnet, we need to determine all the contributions to
the torque acting on the switching magnet. Firstly, there is the
spin-transfer torque ${\bf T}^{\rm s-t}$ which we discussed in
Section \ref{due}. Secondly, there is a torque due to the uniaxial
in-plane and easy plane (shape) anisotropies. The easy-plane shape
anisotropy torque arises because the switching magnet is a thin
layer typically only a few nanometers thick. The in-plane uniaxial
anisotropy is usually also a shape anisotropy arising from an
elongated cross section of the switching magnet \cite{albert}. We
take the uniaxial anisotropy axis of the switching magnet to be
parallel to the $z$-axis of the coordinate system shown in Fig.
\ref{fig1}. Since the switching magnet lies in the $xz$-plane, we
can write the total anisotropy field as
\begin{equation}
\label{hh}
{\bf H}_{\rm A}={\bf H}_{\rm u}+{\bf H}_{\rm p}
\end{equation}
where ${\bf H}_{\rm u}$ and  ${\bf H}_{\rm p}$ are given by
\begin{equation}
\label{hu}
{\bf H}_{\rm u}= H_{\rm u0}({\bf m}\cdot{\bf e}_{z}){\bf e}_{z},
\end{equation}

\begin{equation}
\label{hp}
{\bf H}_{\rm p}=-H_{\rm p0}({\bf m}\cdot{\bf e}_{y}){\bf e}_{y}.
\end{equation}
Here ${\bf e}_x$, ${\bf e}_y$ and ${\bf e}_z$ are unit vectors in
the directions of the axes shown in Fig. 1. If we write the energy
of the switching magnet in the anisotropy field as $-{\bf H}_{\rm
A}\cdot\langle{\bf S}_{\rm tot}\rangle$, where $\langle{\bf S}_{\rm
tot}\rangle$ is the total spin angular momentum of the switching
magnet, them $H_{\rm u0}$,$H_{\rm p0}$ which measure the strengths
of the uniaxial and easy-plane anisotropies have dimensions of
frequency. These quantities may be converted to a field in tesla by
multiplying by $\hbar/2\mu_{\rm
  B}=5.69\times 10^{-12}$.

We are now ready to study the time evolution of the unit vector ${\bf
  m}$ in the direction of the switching magnet moment. The LLG
  equation takes the usual form
\begin{equation}
\label{llg}
\frac{{\rm d}{\bf m}}{{\rm d}t}+\gamma{\bf m}\times\frac{{\rm d}{\bf m}}{{\rm d}t}={\bm\Gamma}
\end{equation}
where the reduced total torque ${\bm\Gamma}$ acting on the switching magnet
is given by
\begin{equation}
\label{Gamma}
{\bm\Gamma}=\left[-\left({\bf H}_{\rm A}+{\bf H}_{\rm
    ext}\right)\times\langle{\bf S}_{\rm tot}\rangle+{\bf
    T}_{\perp}+{\bf T}_{\parallel}\right]/
\left|\langle {\bf S}_{\rm tot}\rangle\right|.
\end{equation}
Here ${\bf H}_{\rm ext}$ is an external field, in the same frequency
units as ${\bf H}_{\rm A}$, and $\gamma$ is the Gilbert damping
parameter. Following Sun \cite{sun}, Eq. (\ref{llg}) may be written
more conveniently as
\begin{equation}
\label{Gamma2}
\left(1+\gamma^2\right)\frac{{\rm d}{\bf m}}{{\rm
    d}t}={\bm\Gamma}-\gamma{\bf m}\times{\bm\Gamma}.
\end{equation}
It is also useful to measure the strengths of all the torques in units
of the strength of the uniaxial anisotropy \cite{sun}. We shall,
therefore, write the total reduced torque ${\bm\Gamma}$ in the form
\begin{equation}
\label{Gamma3}
{\bm\Gamma}=H_{uo}\left\{({\bf m}\cdot{\bf e}_{z}){\bf m}\times{\bf e}_z-
h_{\rm p}({\bf m}\cdot{\bf e}_{y}){\bf m}\times
{\bf  e}_y+v_{\parallel}(\psi)
{\bf m}\times({\bf p}\times{\bf m})+\left[v_{\perp}(\psi)+h_{\rm ext}\right]
{\bf m}\times{\bf p}\right\}
\end{equation}
where the relative strength of the easy plane anisotropy $h_{\rm
  p}=H_{\rm p0}/H_{\rm u0}$ and
  $v_{\parallel}(\psi)=vg_{\parallel}(\psi)$,
$v_{\perp}(\psi)=vg_{\perp}(\psi)$ measure the strengths of the
  torques ${\bf T}_{\parallel}$ and ${\bf T}_{\perp}$. The reduced
  bias is defined by $v=eV_{\rm b}/(|\langle {\bf S}_{\rm
  tot}\rangle|H_{\rm u0})$ and has the opposite sign from the bias
  voltage since $e$ is negative. Thus positive $v$ implies a flow of
  electrons from the polarising to the switching magnet.
The last contribution to the torque in Eq. (\ref{Gamma3}) is due to the
  external field $H_{\rm ext}$ with $h_{\rm ext}=H_{\rm ext}/H_{\rm
  u0}$. The external field is taken in the direction of the
  magnetisation of the polarising magnet, as is the case in most
  experimental  situations.

It follows from Eq. (\ref{llg}) that in a steady state ${\bm\Gamma}=0$. We
shall first consider some cases of experimental importance where the
steady state solutions are trivial and the important physics is
concerned entirely with their stability. To discuss stability, we
linearise Eq. (\ref{Gamma2}), using Eq. (\ref{Gamma3}), about a steady
state solution ${\bf m}={\bf m}_{0}$. Thus
\begin{equation}
\label{Gamma4}
{\bf m}={\bf m}_0+\xi{\bf e}_{\alpha}+\eta{\bf e}_{\phi},
\end{equation}
where ${\bf e}_{\alpha}$, ${\bf e}_{\phi}$ are unit vectors in the
direction ${\bf m}$ moves when $\alpha$ and $\phi$ are increased
independently. The linearised equation may be written in the form
\begin{equation}
\label{Gamma5}
\frac{{\rm d}\xi}{{\rm d}\tau}=A\xi+B\eta;\quad
\frac{{\rm d}\eta}{{\rm d}\tau}=C\xi+D\eta.
\end{equation}
Following Sun \cite{sun}, we have introduced the natural
dimensionless time variable $\tau=tH_{\rm u0}/(1+\gamma^2)$. The
conditions for the steady state to be stable are
\begin{equation}
\label{Gamma6}
F=A+D\leqslant 0;\quad G=AD-BC\geqslant 0
\end{equation}
excluding $F=G=0$ \cite{20}. For simplicity we give these conditions
explicitly only for the case where either
$v_{\parallel}^{\prime}(\psi_0)=v_{\perp}^{\prime}(\psi_0)=0$, with
$\psi_0=\cos^{-1}({\bf p}\cdot{\bf m}_0)$, or ${\bf m}_0=\pm{\bf p}$.
The case ${\bf m}_0=\pm{\bf p}$ is very common experimentally as is
discussed below. The stability condition $G\geqslant 0$ may be written
\begin{eqnarray}
\label{stab}
&&Q^2v_{\parallel}^2+(Qh+\cos 2\alpha_0)(Qh+\cos^2\alpha_0)+h_{\rm
  p}\left\{Qh(1-3\sin^2\phi_0\sin^2\alpha_0)+\cos 2\alpha_0
(1-2\sin^2\alpha_0\sin^2\phi_0)\right\}-\nonumber\\
&&h_{\rm
  p}^2\sin^2\alpha_0\sin^2\phi_0(1-2\sin^2\phi_0\sin^2\alpha_0)
\geqslant  0,
\end{eqnarray}
 where $v_{\parallel}=v_{\parallel}(\psi_0)$,
 $h=v_{\perp}(\psi_0)+h_{\rm ext}$ and $Q=\cos\psi_0$. The condition
 $F\leqslant 0$ takes the form
\begin{equation}
\label{F0}
-2(v_{\parallel}+\gamma h)Q-\gamma(\cos
 2\alpha_0+\cos^2\alpha_0)-\gamma h_{\rm
 p}(1-3\sin^2\phi_0\sin^2\alpha_0)
\leqslant 0.
\end{equation}

We now discuss several interesting examples, the first of these
relating to experiments of Grollier {\it et al.} \cite{18} and
others. In these experiments the magnetisation of the polarising
magnet, the uniaxial anisotropy axis and the external field are all
collinear (along the in-plane $z$-axis in our convention). In this
case the equation ${\bm\Gamma}=0$, with ${\bm\Gamma}$ given by
Eq. (\ref{Gamma3}), shows immediately that possible steady states are
given by ${\bf m}_0=\pm{\bf p}(\alpha_0=0,\pi)$, corresponding to the
switching magnet moment along the $z$-axis. These are the only
solutions when $h_{\rm p}=0$. For $h_{\rm p}\neq 0$ other steady-state
solutions may exist but in the parameter regime which has been
investigated they are always unstable \cite{13}. We shall assume this
is always the case and concentrate on the solutions ${\bf m}_0=\pm{\bf
  p}$. In the state of parallel magnetisation (P) ${\bf m}_0={\bf p}$
we have $v_{\parallel}=vg_{\parallel}(0)$, $h=vg_{\perp}(0)+h_{\rm
  ext}$, $\alpha_0=0$ and $Q=1$. The stability conditions (\ref{stab})
and (\ref{F0}) become
\begin{equation}
\label{stab2}
\left[g_{\parallel}(0)\right]^2v^2+(vg_{\perp}(0)+h_{\rm
  ext}+1)^2+h_{\rm p}\left[vg_{\perp}(0)+h_{\rm
    ext}+1\right]\geqslant 0
\end{equation}

\begin{equation}
\label{F0-2}
g_{\parallel}(0)v+\gamma\left[vg_{\perp}(0)+h_{\rm
  ext}+1+\frac12 h_{\rm p}\right]\geqslant 0.
\end{equation}
In the state of anti-parallel magnetisation (AP) ${\bf m}_0=-{\bf p}$ we
have $v_{\parallel}=vg_{\parallel}(\pi)$, $h=vg_{\perp}(\pi)+h_{\rm
  ext}$, $\alpha_0=\pi$ and $Q=-1$. The stability conditions for the
AP state are thus
\begin{equation}
\label{stab2AP}
\left[g_{\parallel}(\pi)\right]^2v^2+(-vg_{\perp}(\pi)-h_{\rm
  ext}+1)^2+h_{\rm p}\left[-vg_{\perp}(\pi)-h_{\rm
    ext}+1\right]\geqslant 0
\end{equation}

\begin{equation}
\label{F0-2AP}
g_{\parallel}(\pi)v+\gamma\left[vg_{\perp}(\pi)+h_{\rm
  ext}-1-\frac12 h_{\rm p}\right]\leqslant 0.
\end{equation}

In the regime of low external field ($h_{\rm ext}\approx 1$, {\it
  i.e.} $H_{\rm ext}\approx H_{u0}$) we have $H_{\rm p}>>H_{\rm ext}$
  ($h_{\rm p}\approx 100$). Eqs. (\ref{stab2}) and (\ref{stab2AP}) may be
  then approximated by
\begin{equation}
\label{stab2AP-approx}
vg_{\perp}(0)+h_{\rm ext}+1>0
\end{equation}

\begin{equation}
\label{F0-2AP-approx}
vg_{\perp}(\pi)+h_{\rm ext}-1<0.
\end{equation}
Equation (\ref{stab2AP-approx}) corresponds to P stability and
(\ref{F0-2AP-approx}) to AP stability. It is convenient to define
scalar quantities $T_{\perp}$, $T_{\parallel}$ by
$T_{\perp}=g_{\perp}(\psi)\sin\psi$,
$T_{\parallel}=g_{\parallel}(\psi)\sin\psi$, these being scalar
components of spin-transfer torque in units of $eV_{\rm b}$
(cf. Eq. (\ref{torque2})). Then $g_i(0)=[{\rm d}T_i/{\rm
    d}\psi]_{\psi=0}$ and $g_i(\pi)=-[{\rm d}T_i/{\rm
    d}\psi]_{\psi=\pi}$ with $i=\perp,\parallel$. Model calculations
\cite{13} show that both $g_{\perp}$ and $g_{\parallel}$ can be of
either sign, although positive values are more common. Also there is no
general rule about the relative magnitude of $g_i(0)$ and $g_i(\pi)$.

\begin{figure}
\includegraphics[width=13cm]{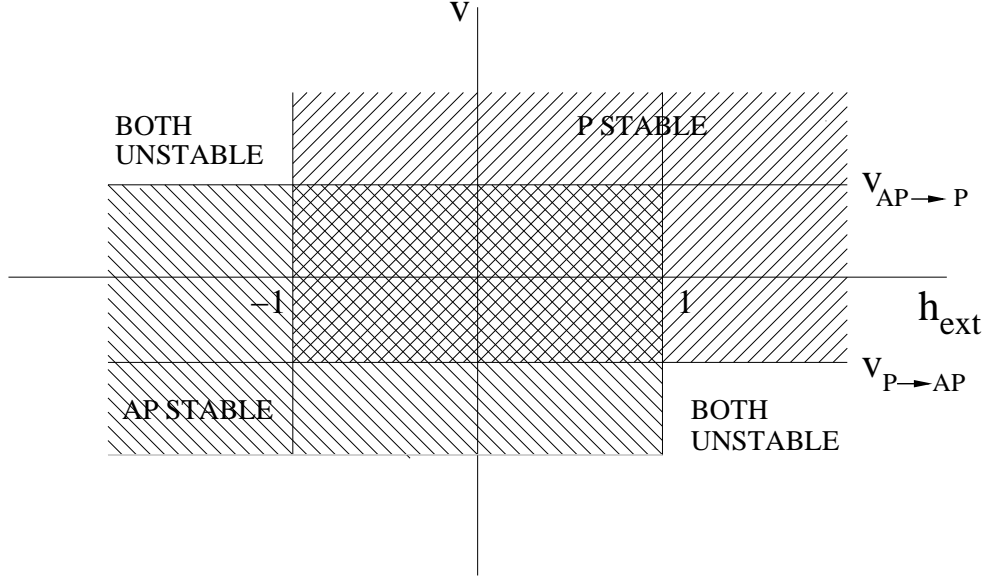}
\caption{Bias-field stability diagram for $g_{\perp}(\psi)=0$,
  $g_{\parallel}(\psi)>0$. A small downward slope of the lines $V_{{\rm
  AP}\rightarrow{\rm P}}$,$V_{{\rm P}\rightarrow{\rm AP}}$
 (see Eq. (\ref{crbias})) is not shown.}
\label{fig4}
\end{figure}

We now illustrate the consequences of the above stability conditions
by considering two limiting cases. We first consider the case
$g_{\perp}(\psi)=0$, $g_{\parallel}>0$, as assumed by Grollier {\it
  et. al} \cite{18} in the analysis of their data. In Fig. 4 we
plot the regions of P and AP stability deduced from
Eqs. (\ref{F0-2}),(\ref{F0-2AP})-(\ref{F0-2AP-approx}), in the
($v$,$h_{\rm ext}$)-plane. Grollier {\it et al.} plot current instead
of bias but this should not change the form of the figure. Theirs is
rather more complicated, owing to a less transparent stability
analysis with unnecessary approximation. The only approximations made
above, to obtain Eqs. (\ref{stab2AP-approx}) and(\ref{F0-2AP-approx}), can
easily be removed, which results in the critical field lines $h_{\rm
  ext}=\pm1$ acquiring a very slight curvature given by $h_{\rm
  ext}\approx 1+[vg_{\parallel}(\pi)]^2/h_{\rm p}$ and $h_{\rm
  ext}\approx -1-[vg_{\parallel}(0)]^2/h_{\rm p}$. The critical biases
in the figure are give by
\begin{eqnarray}
\label{crbias}
v_{{\rm AP}\rightarrow{\rm P}}&=&\gamma \left[1+\frac12 h_{\rm p}-h_{\rm
  ext}\right]/g_{\parallel}(\pi)\nonumber\\
v_{{\rm P}\rightarrow{\rm AP}}&=&-\gamma \left[1+\frac12 h_{\rm p}+h_{\rm
  ext}\right]/g_{\parallel}(0).
\end{eqnarray}
A downward slope from left to right of the corresponding lines in
Fig. \ref{fig4} is not shown there. Since the damping parameter
$\gamma$ is small ($\gamma\approx 0.01$) this downward slope of the
critical bias lines is also small. From Fig. \ref{fig4} we can deduce
the behaviour of resistance versus bias in the external field
regimes $|h_{\rm ext}|<1$ and $|h_{\rm ext}|>1$.

\begin{figure}
\includegraphics[width=13cm]{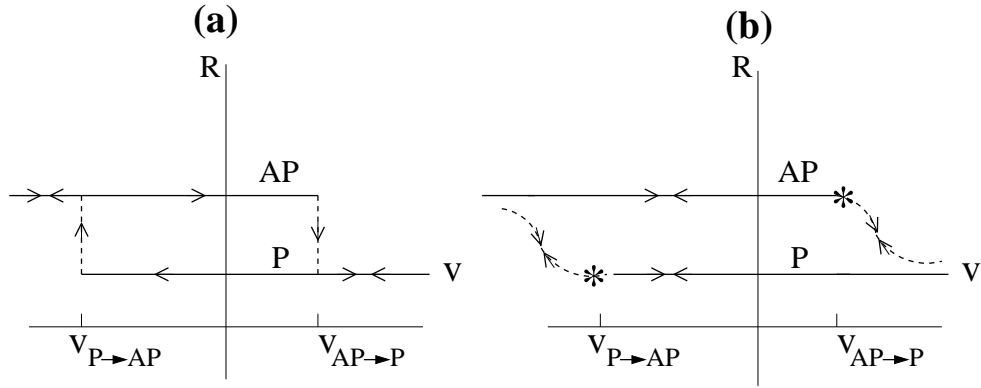}
\caption{(a) Hysteresis loop of resistance vs bias for $|h_{\rm
    ext}|<1$; (b) Reversible behaviour (no hysteresis) for $|h_{\rm
    ext}|<-1$ (upper curve) and $h_{\rm ext}>1$ (lower curve). The
    dashed lines represent hypothetical behaviour of average
    resistance in regions of Fig. \ref{fig4} marked ``both unstable''
    where no steady states exist.}
\label{fig5}
\end{figure}

Consider first the case $|h_{\rm ext}|<1$. Suppose we start in the AP
state with a bias $v=0$ which is gradually increased to $v_{{\rm
    AP}\rightarrow P}$. At this point the AP state becomes unstable
and the system switches to the P state as $v$ increases further. On
reducing $v$ the hysteresis loop is completed via a switch back to
the AP state at the negative bias $v_{{\rm P}\rightarrow AP}$. The
hysteresis loop is shown in Fig. \ref{fig5}(a). The increase in resistance
R between the P and AP states is the same as would be
produced by varying the applied field in a GMR experiment.
Now consider the case $h_{\rm ext}<-1$. Starting again in the AP
state at $v=0$ we see from Fig. \ref{fig4} that, on increasing $v$
to $v_{{\rm AP}\rightarrow{\rm P}}$, the AP state becomes unstable but there is
no stable P state to switch to. This point is marked by an asterisk
in Fig. \ref{fig5}(b). For $v>v_{{\rm AP}\rightarrow{\rm P}}$, 
the moment of the
switching magnet is in a persistently time-dependent state. However,
if $v$ is now decreased below $v_{{\rm P}\rightarrow{\rm AP}}$
the system homes in on the stable AP state and the overall behaviour 
is reversible,
{\it i.e.} no switching and no hysteresis occur. When $h_{\rm
ext}>1$ similar behaviour, now involving the P state, occurs at
negative bias, as shown in Fig. \ref{fig5}(b). The dashed curves in
Fig. \ref{fig5}(b) show a hypothetical time-averaged resistance in the
regions of time-dependent magnetisation. As discussed later
time-resolved measurements of resistance suggest that several
different types of dynamics can occur in these regions.

It is clear from Fig. \ref{fig5}(a) that the jump AP$\rightarrow$P
always occurs for positive bias $v$, which corresponds to flow of
electrons from the polarising to the switching magnet. This result
depends on the assumption that $g_{\parallel}>0$; if
$g_{\parallel}<0$ it is easy
  to see that the sense of the hysteresis loop is reversed and the
  jump P$\rightarrow$AP occurs for positive $v$. To our knowledge this
  reverse jump has never been observed, although $g_{\parallel}<0$ can
  occur in principle and is predicted theoretically \cite{13} for the
  Co/Cu/Co(111) system with a switching magnet consisting of a single
  atomic plane of Co. It follows from Eq. (\ref{crbias}) that
$|v_{{\rm P}\rightarrow{\rm AP}}/v_{{\rm AP}\rightarrow{\rm
    P}}|=|g_{\parallel}(\pi)/g_{\parallel}(0)|$ in zero external
field. Experimentally this ratio, essentially the same as the ratio of
critical currents, may be considerably less than 1 ({\it e.g.} $<0.5$
\cite{albert}), greater than 1 ({\it e.g.} $\approx 2$ \cite{19}) or
close to 1 \cite{16}. Usually the field dependence of the critical
current is found to be stronger than that predicted by Eq. (\ref{crbias})
\cite{albert,16}.

We now discuss the reversible behaviour shown in Fig. \ref{fig5}(b)
which occurs for $|h_{\rm ext}|>1$. The transition from hysteretic
to reversible behaviour at a critical external field seems to have
been first seen in pillar structures by Katine {\it et al.}
\cite{21}. Curves similar to the lower one in Fig. \ref{fig5}(b) are
reported with $|v_{{\rm P}\rightarrow{\rm AP}}|$ increasing with
increasing $h_{\rm ext}$, as expected from Eq. (\ref{crbias}). Plots
of the differential resistance ${\rm d}V/{\rm d}I$ show a peak near
the point of maximum gradient of the dashed curve. Similar
behaviour has been reported by several groups \cite{22,23,24}. It
is particularly clear in the work of Kiselev {\it at al.} \cite{22}
that the transition from hysteretic behaviour (as in Fig.
\ref{fig5}(a)) to reversible behaviour with peaks in ${\rm d}V/{\rm
d}I$ occurs at the coercive field 600 Oe of the switching layer
($h_{\rm ext}=1$). The important point about the peaks in ${\rm
d}V/{\rm d}I$ is that for a given sign of $h_{\rm ext}$ they only
occur for one sign of the bias. This clearly shows that this effect
is due to spin-transfer and not to Oersted fields. Myers {\it et
al.} \cite{25} show a current-field stability diagram similar to the
bias-field one of Fig. \ref{fig4} with a critical field of 1500 Oe.
They examine the time dependence of the resistance at room
temperature with the field and current adjusted so that the system
is in the ``both unstable'' region in the fourth quadrant of Fig.
\ref{fig4} but very close to its top left-hand corner. They observe
telegraph-noise-type switching between approximately P and AP
states with slow switching times in the range 0.1-10 s. Similar
telegraph noise with faster switching times was observed by Urazhdin
{\it et al.} \cite{23} at current and field close to a peak in ${\rm
d}V/{\rm d}I$. In the region of P and AP instability Kiselev {\it et
al.} \cite{22} and Pufall {\it et al.} \cite{24} report various
types of dynamics of precessional type and random telegraph
switching type in the microwave Ghz regime. Kiselev {\it et al.}
\cite{22} propose that systems of the sort considered here might
serve as nanoscale microwave sources or oscillators, tunable by
current and field over a wide frequency range.

\begin{figure}
\includegraphics[width=10cm]{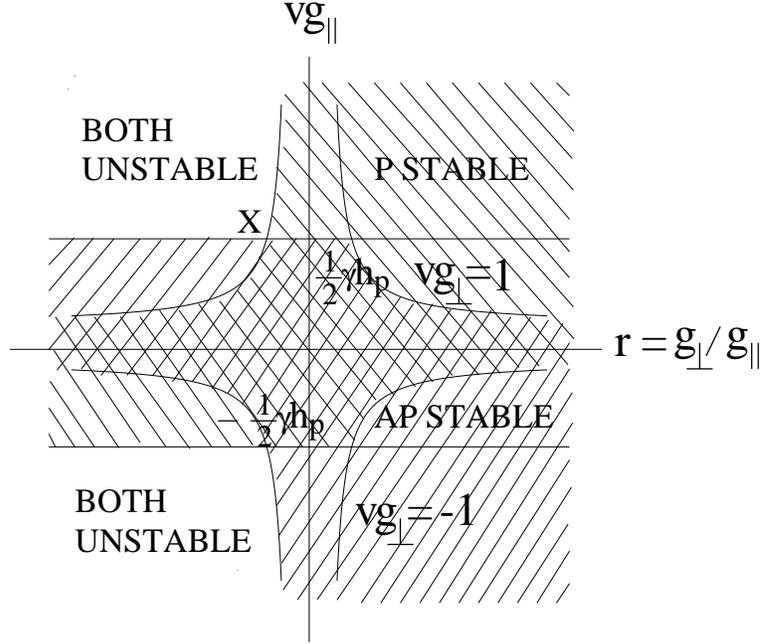}
\caption{Stability diagram for $h_{\rm ext}=0$.} \label{fig6}
\end{figure}

\begin{figure}
\includegraphics[width=13cm]{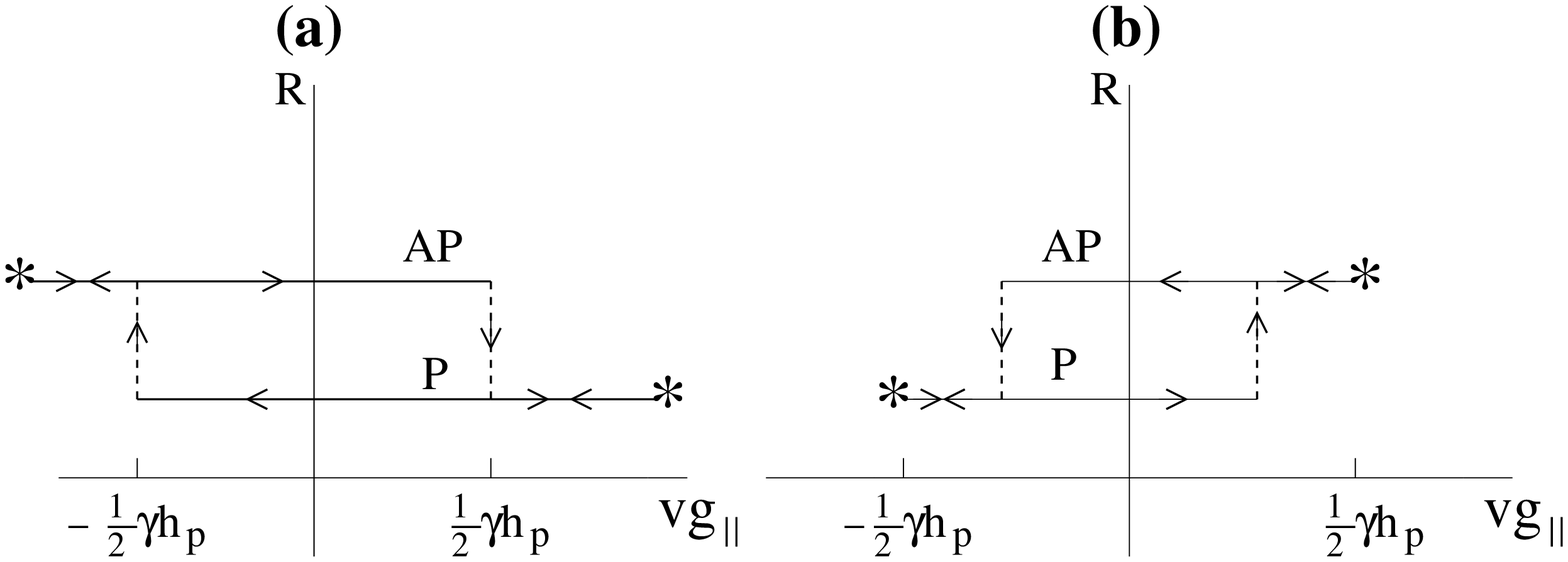}
\caption{Hysteresis loop for (a) $r_{\rm c}<r<0$; (b) $r<r_{\rm
c}$.} \label{fig7}
\end{figure}

We now return to the stability conditions
(\ref{F0-2}),(\ref{F0-2AP})-(\ref{F0-2AP-approx}) and consider the case
of $g_{\perp}(\psi)\neq 0$ but $h_{\rm ext}=0$. These conditions of
stability of the P state may be written approximately, remembering
that $\gamma << 1$, $h_{\rm p}>> 1$, as
\begin{equation}
\label{vgperp}
vg_{\perp}(0)>-1, \quad vg_{\parallel}(0)>-\frac12 \gamma h_{\rm p}.
\end{equation}
The conditions for stability of the AP state are
\begin{equation}
\label{vgperpAP}
vg_{\perp}(\pi)<1, \quad vg_{\parallel}(\pi)<\frac12 \gamma h_{\rm p}.
\end{equation}
In Fig. \ref{fig6} we plot the regions of P and AP stability,
assuming $g_{\perp}(0)=g_{\perp}(\pi)=g_{\perp}$ and
$g_{\parallel}(0)=g_{\parallel}(\pi)=g_{\parallel}$ for simplicity.
We also put $r=g_{\perp}/g_{\parallel}$.
For $r>0$ we find the normal hysteresis loop as in Fig.
\ref{fig5}(a) if we plot $R$ against $vg_{\parallel}$ (valid for
either sign of $g_{\parallel}$). In Fig. \ref{fig7} we plot the
hysteresis loops for the cases $r_{\rm c}<r<0$ and $r<r_{\rm c}$,
where $r_{\rm c}=-2/(\gamma h_{\rm p})$ is the value of $r$ at the
point $X$ in Fig. \ref{fig6}.
The points labelled by asterisks have the same significance as in
Fig. \ref{fig5}(b). If in Fig. \ref{fig7}(a) we increase
$vg_{\parallel}$ beyond its value indicated by the right-hand
asterisk we move into the ``both-unstable'' region where the
magnetisation direction of the switching magnet is perpetually in a
time-dependent state. Thus negative $r$ introduces behaviour in zero
applied field which is similar to that found when the applied field
exceeds the coercive field of the switching magnet for $r=0$. This
behaviour was predicted by Edwards {\it et al.} \cite{13}, in
particular for a Co/Cu/Co(111) system with the switching magnet
consisting of a Co monolayer. Zimmler {\it et al.} \cite{26} use
methods similar to the ones described here to analyse their data on
a Co/Cu/Co nanopillar and deduce that $g_{\parallel}>0$,
$r=g_{\perp}/g_{\parallel}\approx -0.2$. It would be interesting to
carry out time-resolved resistance measurements on this system at
large current density (corresponding to $vg_{\perp}<-1$) and zero
external field. 

So far we have considered the low-field regime
($H_{\rm ext}\approx$ coercive field of switching magnet) with both
magnetisations and the external field in-plane. There is another
class of experiments in which a high field, greater than the
demagnetising field ($>2T$), is applied perpendicular to the plane
of the layers. The magnetisation of the polarising magnet is then
also perpendicular to the plane. This is the situation in the early
experiments where a point contact was employed to inject high
current densities into magnetic multilayers \cite{27,28,29}. In this
high-field regime a peak in the differential resistance ${\rm
d}V/{\rm d}I$ at a critical current was interpreted as the onset of
current-induced excitation of spin waves in which the spin-transfer
torque leads to uniform precession of the magnetisation
\cite{6,27,28}. No hysteretic magnetisation reversal was observed
and it seemed that the effect of spin-polarised current on the
magnetisation is quite different in the low- and high-field regimes.
Recently, however, \"{O}zyilmaz {\it et al.} \cite{30} have studied
Co/Cu/Co nanopillars ($\approx 100$nm in diameter) at $T=4.2$K for
large applied fields perpendicular to the layers. They observe
hysteretic magnetisation reversal and interpret their results using
the Landau-Lifshitz equation. We now give a similar discussion
within the framework of this section.

\begin{figure}
\includegraphics[width=11cm]{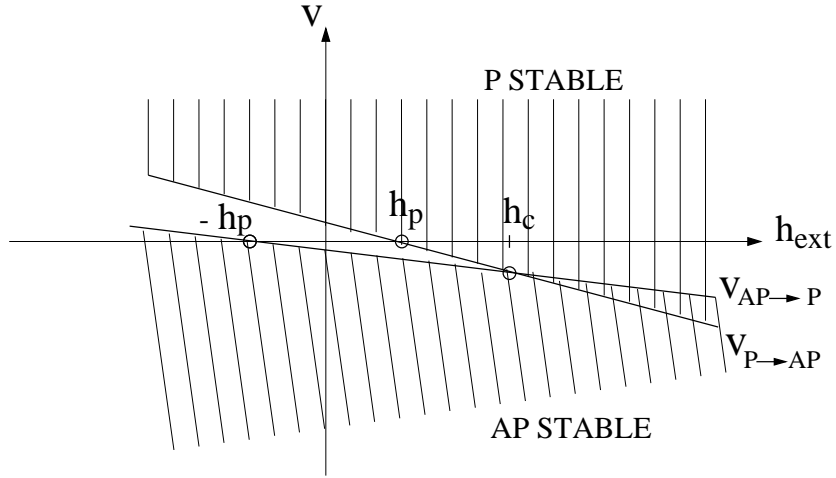}
\caption{Bias-field stability diagram for large external field
($h_{\rm ext}>h_{\rm p}$) perpendicular to the layers. }
 \label{fig8}
\end{figure}

Following \"Ozyilmaz {\it et al.}, we neglect the uniaxial
anisotropy term in Eq. (\ref{Gamma3}) for the reduced torque ${\bm\Gamma}$
while retaining $H_{u0}$ as a scalar factor. Hence
\begin{equation}
\label{gamma25}
{\bm\Gamma}=H_{u0}\left\{\left[h_{\rm
ext}+v_{\perp}(\psi)-h_{\rm
p}\cos\psi\right]{\bf m}\times{\bf p}+v_{\parallel}(\psi){\bf m}\times({\bf p}\times{\bf m})\right\}
\end{equation}
where ${\bf p}$ is the unit vector perpendicular to the plane. When
$v_{\parallel}(\psi)\neq 0$ the only possible steady-state solutions
of ${\bm\Gamma} =0$ are ${\bf m}_0=\pm{\bf p}$. On linearizing Eq. \ref{Gamma2}
about ${\bf m}_0$ as before we find that the condition $G\geqslant
0$ is always satisfied. The second stability condition $F<0$ becomes
\begin{equation}
\label{stabcond} 
\left[v_{\parallel}(\psi_0)+\gamma(v_{\perp}(\psi_0)+h_{\rm
ext}-h_{\rm p})\right]\cos\psi_0>0
\end{equation}
where $\psi_0=\cos^{-1}({\bf m}_0\cdot{\bf p})$. Applying this to the P
state ($\psi_0=0$) and the AP state ($\psi_0=\pi$) we obtain the
conditions
\begin{equation}
\label{conditiona} v>\gamma(h_{\rm p}-h_{\rm ext})/g(0)
\end{equation}
\begin{equation}\label{conditionb}
v<-\gamma(h_{\rm p}+h_{\rm ext})/g(\pi), 
\end{equation}
where the first condition applies to the P stability and the second
to the AP stability. Here $g(\psi)=g_{\parallel}(\psi)+\gamma
g_{\perp}(\psi)$. The corresponding stability diagram is shown in
Fig. \ref{fig8}, where we have assumed $g(\pi)>g(0)>0$ for
definiteness.
The boundary lines cross at $h_{\rm ext}=h_{\rm c}$, where $h_{\rm
c}=h_{\rm p}[g(\pi)+g(0)]/[g(\pi)-g(0)]$. This analysis is only
valid for fields larger than the demagnetising field ($h_{\rm
ext}>h_{\rm p}$) and we see from the figure that for $h_{\rm
ext}>h_{\rm c}$ hysteretic switching occurs. This takes place for
only one sign of the bias (current) and the critical biases
(currents) increase linearly with $h_{\rm ext}$ as does the width of
the hysteresis loop $|v_{{\rm P}\rightarrow{\rm AP}}-v_{{\rm
AP}\rightarrow{\rm P}}|$. This accords with the observations of
\"Ozyilmaz {\it et al.} The critical currents are not larger than
those in the low-field or zero-field regimes (cf. Eqs.
(\ref{conditiona}), (\ref{conditionb}) with Eq. (\ref{crbias})) and yet the
magnetisation of the switching magnet can be switched against a very
large external field. However, in this case the AP state is only
stabilised by maintaining the current.

The experiments on spin transfer discussed above have mainly been
carried out at constant temperature, typically $4.2$K or room
temperature. The effect on current-driven switching of varying the
temperature has recently been studied by several groups
\cite{23,25,31}. The standard N\'eel-Brown theory of thermal
switching \cite{32} does not apply because the Slonczewski in-plane
torque is not derivable from an energy function. Li and Zhang
\cite{33} have generalised the standard stochastic Landau-Lifschitz
equation, which includes white noise in the effective applied field,
to include spin transfer torque. In this way they have successfully
interpreted some of the experimental data. A full discussion of this
work is outside the scope of the present review. However it should
be pointed out that in addition to the classical effect of white
noise there is an intrinsic temperature dependence of quantum
origin. This arises from the Fermi distribution functions which
appear in expressions for the spin-transfer torque (see Eqs.
(\ref{torque}) and (\ref{landau33})).

So far we have discussed steady-state solutions of the LLG equation
(\ref{Gamma2}). It is important to study the magnetisation dynamics
of the switching layer in the situation during the jumps
AP$\rightarrow$P and P$\rightarrow$AP of the hysteresis curve in
zero external field, and secondly under conditions where only
time-dependent solutions are possible, for example in the regions of
sufficiently strong  current and external field marked ''both
unstable" in Fig. \ref{fig4}. The first situation has been studied
by Sun \cite{sun}, assuming single-domain behaviour of the switching
magnet, and by Miltat {\it et al.} \cite{34} with more general
micromagnetic configurations. Both situations have been considered
by Li and Zhang \cite{35}. In the second case they find precessional
states, and the possibility of ''telegraph noise" at room
temperature, as seen experimentally in Refs. \cite{22,24}. Switching
times (AP$\rightarrow$P and P$\rightarrow$AP) are estimated to be of
the order 1ns. Micromagnetic simulations \cite{34} indicate that the
Oersted field cannot be completely ignored for typical pillars with
diameter of the order of 100nm.

Finally, in this section, we briefly discuss some practical
considerations which may ultimately decide whether current-induced
switching is useful in spintronics. Sharp switching, with nearly
rectangular hysteresis loops, is obviously desirable and this
demands single-domain behaviour. In experiments on nanopillars of
circular cross section \cite{21} multidomain behaviour was observed
with the switching transition spread over a range of current.
Subsequently the same group \cite{albert} found sharp switching in
pillars whose cross-section was an elongated hexagon, which
introduces strong uniaxial in-plane shape anisotropy. It was known
from earlier magnetisation studies of nanomagnet arrays \cite{36}
that such a shape anisotropy can result in single domain behaviour.
A complex switching transition need not necessarily indicate
multidomain behaviour. It could also arise from a marked departure
of $T_{\perp}(\psi)$ and/or $T_{\parallel}(\psi)$ from sinusoidal
behaviour, such as occurs near $\psi=\pi$ in calculations for
Co/Cu/Co(111) with two atomic planes of Co in the switching magnet
(see Fig. \ref{fig13}(b)). In the calculations of the corresponding
hysteresis loops (Fig. \ref{fig16}) the torques were approximated by
sine curves but an accurate treatment would certainly complicate the
AP$\rightarrow$P transition which occurs at negative bias in Fig.
\ref{fig16}(b). Studies of this effect are planned.

The critical current density for switching is clearly an important
parameter. From Eq. (\ref{crbias}) the critical reduced bias for the
P$\rightarrow$AP transition is to a good approximation given by
$-\gamma h_{\rm p}/[2g_{\parallel}(0)]$. Using the definitions of
reduced quantities given after Eq. (\ref{Gamma3}), we may write the
actual critical bias in volts as
\begin{equation} \label{VPAP}
V_{P\rightarrow AP}=M\gamma M_{\rm s}H_{\rm d}/[2g_{\parallel}(0)|e|],
\end{equation}
where $M$ is the number of atomic planes in the switching magnet,
$M_{\rm s}$ is the average moment ($J/T$) of the switching magnet
per atomic plane per unit area, and $H_{\rm d}=\hbar H_{\rm
p0}/(2\mu_{\rm B})$ is the easy-plane anisotropy field in tesla. As
expressed earlier $g_{\parallel}(0)=({\rm d}T_{\parallel}/{\rm
d}\psi)_{\psi =0}$ where the torque $T_{\parallel}$ is per unit area
in units of $eV_{\rm B}$. (The calculated torques in Figs.
\ref{fig13} and \ref{fig14} of Sec. \ref{cinque} are per surface
atom so that if these are used to determine $g_{\parallel}(0)$ in
Eq. (\ref{VPAP}) $M_{\rm s}$ must be taken per surface atom.)

An obvious way to reduce the critical bias, and hence the critical
current, is to reduce $M$, the thickness of the switching magnet.
Calculations show \cite{13} (see also Fig. \ref{fig14}) that
$g_{\parallel}$ does not decrease with $M$ and may, in fact,
increase for small values such as $M=2$. Careful design of the
device might also increase $g_{\parallel}(0)$ beyond the values
($<0.01$ per surface atom) which seem to be obtainable in simple
trilayers \cite{13}. Jiang {\it et al.} \cite{37,38}, have studied
various structures in which the polarising magnet is pinned by an
adjacent antiferromagnet (exchange biasing) and in which a thin Ru
layer is incorporated between the switching layer and the lead.
Critical current densities of $2\times 10^{6}$Acm$^{-2}$ have been
obtained which are substantially lower than those in Co/Cu/Co
trilayers. Such structures can quite easily be investigated
theoretically by the methods of Section \ref{cinque}.

Decreasing the magnetisation $M_{\rm s}$, and hence the
demagnetising field ($\propto H_{\rm d}$), would be favourable but
$g_{\parallel}$ then tends to decrease also \cite{13}. A possible
way of decreasing $H_{\rm d}$ without decreasing local magnetic
moments in the system is to use a synthetic ferrimagnet as the
switching magnet \cite{39}. The Gilbert damping factor $\gamma$ is
another crucial parameter but it is uncertain whether this can be
decreased significantly. However, the work of Capelle and Gyorffy
\cite{40} is an interesting theoretical development. The search for
structures with critical current densities low enough for use in
spintronic devices ($10^{5}$Acm$^{-2}$ perhaps) \cite{41} is an
enterprise where experiment and quantitative calculations \cite{13}
should complement each other fruitfully.

\section{Quantitative theory of spin-transfer torque}\label{quattro}
\subsection{General principles}

To put the phenomenological treatment of Sec. \ref{tre} on a
first-principle quantitative basis we must calculate the
spin-transfer torques (Eqs. (\ref{torque2}) in a steady state for
real systems. For this purpose it is convenient to describe the
magnetic and nonmagnetic layers of Fig. \ref{fig1} by tight-binding
models, in general multiorbital with s, p, and d orbitals whose
one-electron parameters are fitted to first-principle bulk band
structure \cite{42}. The hamiltonian is therefore of the form
\begin{equation} \label{HAM}
H=H_0+H_{\rm int}+H_{\rm anis}
\end{equation}
where the one-electron hopping term $H_0$ is given by
\begin{equation} \label{H0}
H_0=\sum_{k_{\parallel}\sigma}\sum_{m\mu,n\nu}t_{m\mu,n\nu}({\bf
k}_{\parallel})
c^{\dagger}_{{\bf k}_{\parallel}m\mu\sigma}
c_{{\bf k}_{\parallel}n\nu\sigma},
\end{equation}
where $c^{\dagger}_{k_{\parallel}m\mu\sigma}$ creates an electron in
a Bloch state, with in-plane wave vector ${\bf k}_{\parallel}$ and
spin $\sigma$, formed from a given atomic orbital $\mu$ in plane
$m$. Eq. \ref{H0} generalises the single orbital eq.
(\ref{hamiltonian}). $H_{\rm int}$ is an on-site interaction between
electrons in d orbitals which leads to an exchange splitting of the
bands in the ferromagnets and is neglected in the spacer and lead.
Finally, $H_{\rm anis}$ contains anisotropy fields in the switching
magnet and is given by
\begin{equation}
\label{Hanis}
H_{\rm anis}=-\sum_{n}{\bf S}_{n}\cdot{\bf H}_{\rm A},
\end{equation}
where ${\bf S}_n$ is the operator of the total spin angular momentum
of plane $n$ and ${\bf H}_{\rm A}$ is given by Eqs. (\ref{hh})-(\ref{hp})
with the unit vector ${\bf m}$ in the direction of
$\sum_{n}\langle{\bf S}_n\rangle$, where $\langle{\bf S}_n\rangle$
is the thermal average of ${\bf S}_n$. We assume here that the
anisotropy fields $H_{\rm u0}$,$H_{\rm p}$ are uniform throughout
the switching magnet but we could generalise to include, for example,
a surface anisotropy.

In the tight-binding description, the spin angular momentum operator
${\bf S}_n$ is given by
\begin{equation}
\label{Sn}
{\bf S}_{n}=\frac12\hbar\sum_{k_{\parallel\mu}}
(c^{\dagger}_{k_{\parallel}n\mu\uparrow},
c^{\dagger}_{k_{\parallel}n\mu\downarrow})
{\bm\sigma}(c_{k_{\parallel}n\mu\uparrow},
c_{k_{\parallel}n\mu\downarrow})^{\rm T}
\end{equation}
and the corresponding operator for the spin angular momentum current
between planes $n-1$ and $n$ is
\begin{equation}
\label{jj}
{\bf j}_{n-1}=-\frac12 \hbar\sum_{{\bf k}_{\parallel}\mu\nu}
t({\bf k}_{\parallel})_{n\nu,n-1\mu}
\left(c^{\dagger}_{k_{\parallel}n\nu\uparrow},
c^{\dagger}_{k_{\parallel}n\nu\downarrow}\right)
{\bm\sigma}
\left(c_{k_{\parallel}n-1\mu\uparrow},
c_{k_{\parallel}n-1\mu\downarrow}\right)^{\rm T}+{\rm h.c.},
\end{equation}
which generalises the single orbital expression (\ref{gamma}). The
rate of change of ${\bf S}_n$ in the switching magnet is given by
\begin{equation}\label{rate}
{\rm i}\hbar\frac{{\rm d}{\bf S}_n}{{\rm d}t}=[{\bf S}_n,H_0]+[{\bf
S}_n,H_{\rm anis}].
\end{equation}
This results holds since the spin operator commutes with the
interaction hamiltonian $H_{\rm int}$.

It is straightforward to show that
\begin{equation}
\label{commut}
[{\bf S}_n,H_0]={\rm i}\hbar({\bf j}_{n-1}-{\bf j}_{n}),
\end{equation}
and
\begin{equation}\label{commut2}
[{\bf S}_n,H_{\rm anis}]=-{\rm i}\hbar({\bf H}_{\rm A}\times{\bf
S}_n).
\end{equation}
On taking the thermal average, Eq. (\ref{rate}) becomes
\begin{equation}\label{thermal}
\langle\frac{{\rm d}{\bf S}_n}{{\rm d}t}\rangle=\langle
{\bf j}_{n-1}\rangle-\langle {\bf j}_n\rangle-{\bf H}_{\rm
  A}\times\langle
{\bf S}_{\rm tot}
\rangle,
\end{equation}
This corresponds to an equation of continuity, stating that the rate
of change of spin angular momentum on plane $n$ is equal to the
difference between the rate of flow of this quantity onto and off
the plane, plus the rate of change due to precession around the
field ${\bf H}_{\rm A}$. When Eq. (\ref{thermal}) is summed over all
planes in the switching magnet we have
\begin{equation}\label{thermal2}
\frac{{\rm d}}{{\rm d}t} \langle{\bf S}_{\rm tot}\rangle=
{\bf T}^{\rm s-t}-{\bf H}_{\rm
A}\times \langle {\bf S}_{\rm tot}\rangle,
\end{equation}
 where the total spin-transfer torque ${\bf T}^{\rm s-t}$ is
given by Eq. (\ref{torque}) and $\langle{\bf S}_{\rm tot}\rangle$ is
the total spin angular momentum of the switching magnet. Equation
(\ref{thermal2}) is equivalent to Eq. (\ref{llg}), for zero external
field, in the absence of damping. Equation (\ref{torque}) shows how
${\bf T}^{\rm s-t}$ required for the phenomenological treatment of Sec.
\ref{tre} is to be determined from the calculated spin currents in
the spacer and lead. As discussed in Sec. \ref{tre}, the
magnetization of a single-domain sample is essentially uniform and
the spin-transfer torque ${\bf T}^{\rm s-t}$ depends on the angle $\psi$
between the magnetisations of the polarising and switching magnets.

To consider time-dependent solutions of Eq. (\ref{llg}) it is
necessary to calculate ${\bf T}^{\rm s-t}$ for arbitrary angle $\psi$ and
for this purpose ${\bf H}_{\rm A}$ can be neglected. To reduce the
calculation of the spin-transfer torque to effectively a
one-electron problem, we replace $H_{\rm int}$ by a selfconsistent
exchange field term $-\sum_n{\bf S}_n\cdot{\bm\Delta}_n$, where the
exchange field ${\bm\Delta}_n$ should be determined
selfconsistently in the spirit of an unrestricted Hartree-Fock (HF)
or local spin density (LSD), approximation. The essential
selfconsistency condition in any HF or LSD calculation is that the
local moment $\langle{\bf S}_n\rangle$ in a steady state is in the
same direction as ${\bm\Delta}_n$. Thus we require
\begin{equation}\label{delta}
{\bm\Delta}_n\times\langle{\bf S}_n\rangle=0
\end{equation}
for each atomic plane of the switching magnet. It is useful to
consider first the situation when there is no applied bias and the
polarising and switching magnets are separated by a spacer which is
so thick that the zero-bias oscillatory exchange coupling \cite{44}
is negligible. In that case we have two independent magnets and the
selfconsistent exchange field in every atomic plane of the switching
magnet is parallel to its total magnetisation which is uniform and
assumed to be along the $z$-axis. Referring to Fig. \ref{fig1} the
selfconsistent solution therefore corresponds to uniform exchange
fields in the polarising and switching magnets which are at an
assumed angle $\psi=\theta$ with respect to one another.

When a bias $V_{\rm b}$ is applied , with a uniform exchange field
${\bm\Delta }=\Delta{\bf e}_z$ in the switching magnet imposed, the
calculated local moments $\langle{\bf S}_n\rangle$ will deviate from
the $z$-direction so that the solution is not selfconsistent. To
prepare a selfconsistent state with ${\bm\Delta}$ and all
$\langle{\bf S}_n\rangle=\langle {\bf S}\rangle$ in the
$z$-direction it is necessary to apply fictitious constraining
fields ${\bf H}_n$ of magnitude proportional to $V_{\rm b}$. The
local field for plane  $n$ is thus ${\bm\Delta}+{\bf H}_n$ but to
calculate the spin currents in the spacer and lead, and hence
${\bf T}^{\rm s-t}$ from Eq. (\ref{torque}), the fields ${\bf H}_n$, of
the order of $V_{\rm b}$, may be neglected compared with ${\bf
\Delta}$. Although the fictitious constraining fields ${\bf H}_n$
need therefore never be calculated, it is interesting to see that
they are in fact related to ${\bf T}^{\rm s-t}$. For the constrained
self-consistent steady state ($\langle{\bf S}_n\rangle=\langle {\bf
S}\rangle$, $\langle{\bf \dot{S}}_n\rangle=0$) in the presence of the
constraining fields, with ${\bf H}_{\rm A}$ neglected as discussed
above, it follows from Eq. (\ref{thermal}) that
\begin{equation}
\label{jjj} \langle{\bf j}_{\bf n-1}\rangle-\langle{\bf j}_{\bf
n}\rangle=(\Delta+{\bf H}_n)\times\langle S\rangle={\bf
H}_n\times\langle{\bf S}\rangle,
\end{equation}
where the local field ${\bm\Delta}+{\bf H}_n$ replaces ${\bf
H}_{\rm A}$. On summing over all atomic planes $n$ in the switching
magnet we have 
\begin{equation}
\label{ttt} {\bf T}^{\rm s-t}=\langle{\bf j}_{\rm
spacer}\rangle-\langle{\bf j}_{\rm lead}\rangle=\sum_n{\bf
H}_n\times\langle{\bf S}\rangle.
\end{equation}
Thus, as expected, in the prepared state with a given angle $\psi$
between the magnetisations of the magnetic layers the spin-transfer
torque is balanced by the total torque due to the constraining
fields.

In the simple model of Section \ref{due}, with infinite exchange
splitting in the magnets, the local moment is constrained to be in
the direction of the exchange field so the question of
selfconsistency is not raised.

The main conclusion of this Section is that the spin-transfer torque
for a given angle $\psi$ between magnetisations may be calculated
using uniform exchange fields making the same angle with one
another. Such calculations are described in Sec. \ref{due} and
\ref{cinque}. The use of this spin-transfer torque in the LLG
equation of Section \ref{tre} completes what we shall call the
''standard model" (SM). It underlies the original work of
Slonczewski \cite{3} and most subsequent work. The spin-transfer
torque calculated in this way should be appropriate even for
time-dependent solutions of the LLG equation. This is based on the
reasonable assumption that the time for the electronic system to
attain a ''constrained steady state" with given $\psi$ is short
compared with the time-scale ($\approx$1ns) of the macroscopic
motion of the switching magnet moment.

Although the SM is a satisfactory way of calculating the
spin-transfer torque its lack of selfconsistency leads to some
non-physical concepts. The first of these is the "transverse spin
accumulation" in the switching magnet \cite{46,47}, This refers to
the deviations of local moments $\langle{\bf S}_n\rangle$ from the
direction of the exchange field, assumed uniform in the SM. In a
self-consistent treatment such deviations do not occur because the
exchange field is always in the direction of the local moment. A
related non-physical concept is the ''spin decoherence length" over
which the spin accumulation is supposed to decay \cite{46,47}, More
detailed critiques of these concepts are given elsewhere
\cite{13,em}.

\subsection{Keldysh formalism for fully realistic calculations of the 
spin-transfer torque}

The wave-function approach to spin-transfer torque described in
Section \ref{due} is difficult to apply to realistic multiorbital
systems. For this purpose Green functions are much more convenient
and Keldysh \cite{11} developed a Green function approach to the
non-equilibrium problem of electron transport. In this section we
apply this method to calculate spin currents in a magnetic layer
structure, following Edwards {\it et al.} \cite{13}.

The structure we consider is shown schematically in Fig. \ref{fig1}.
It consists of a thick (semi-infinite) left magnetic layer
(polarising magnet), a nonmagnetic metallic spacer layer of $N$
atomic planes, a thin switching magnet of $M$ atomic planes, and a
semi-infinite lead. The broken line between the atomic planes $n-1$
and $n$ represents a cleavage plane separating the system into two
independent parts so that charge carriers cannot move between the
two surface planes $n-1$ and $n$. It will be seen that our ability
to cleave the whole system in this way is essential for the
implementation of the Keldysh formalism. This can be easily done
with a tight-binding parametrisation of the band structure by simply
switching off the matrix of hopping integrals $t_{n\nu,n-1\mu}$
between atomic orbitals $\nu$, $\mu$ localised in planes $n-1$ and
$n$. We therefore adopt the tight-binding description with the
Hamiltonian defined by Eqs. (\ref{HAM}-\ref{Sn}).

To use the Keldysh formalism \cite{11,12,53} to calculate the
charge or spin currents flowing between the planes $n-1$ and $n$, we
consider an initial state at time $\tau=-\infty$ in which the
hopping integral $t_{n\nu,n-1\mu}$ between planes $n-1$ and $n$ is
switched off. Then both sides of the system are in equilibrium but
with different chemical potentials $\mu_{\rm L}$ on the left and
$\mu_{\rm R}$ on the right, where $\mu_{\rm L}-\mu_{\rm R}=eV_{\rm
b}$. The interplane hopping is then turned on adiabatically and the
system evolves to a steady state. The cleavage plane, across which
the hopping is initially switched off, may be taken in either the
spacer or in one of the magnets or in the lead. In principle, the
Keldysh method is valid for arbitrary bias $V_{\rm b}$ but here we
restrict ourselves to small bias corresponding to linear response.
This is always reasonable for a metallic system. For larger bias,
which might occur with a semiconductor or insulator as spacer,
electrons would be injected into the right part of the system far
above the Fermi level and many-body processes neglected here would
be important. Following Keldysh \cite{11,12}, we define a two-time
matrix
\begin{equation}\label{kel}
G_{\rm RL}^{+}(\tau,\tau^{\prime})={\rm i}\langle c_{\rm
L}^{\dagger}(\tau^{\prime})c_{\rm R}(\tau)\rangle,
\end{equation}
where $R\equiv(n,\nu,\sigma^{\prime})$ and
$L\equiv(n-1,\mu,\sigma)$, and we suppress the $k_{\parallel}$
label. The thermal average in Eq. (\ref{kel}) is calculated for the
steady state of the coupled system. The matrix $G_{\rm
RL}^{\dagger}$ has dimensions $2m\times 2m$ where $m$ is the number
of orbitals on each atomic site, and is written so that the $m\times
m$ upper diagonal block contains matrix elements between $\uparrow$
spin orbitals and the $m\times m$ lower diagonal block relates to
$\downarrow$ spin. $2m\times 2m$ hopping matrices $t_{\rm LR}$ and
$t_{\rm RL}$ are written similarly and in this case only the
diagonal blocks are nonzero. If we denote $t_{\rm LR}$ by $t$, then
$t_{\rm RL}=t^{\dagger}$. We also generalise the definition of
${\bm\sigma}$ 
so that its components are now direct products of the
$2\times 2$ Pauli matrices $\sigma_x$, $\sigma_y$, $\sigma_z$, and
the $m\times m$ unit matrix. The thermal average of the spin current
operator, given by Eq. (\ref{rate}), may now be expressed as
\begin{equation}\label{uffa}
\langle{\bf j}_{n-1}\rangle=\frac12\sum_{{\bf k}_{\parallel}}{\rm
Tr}\left\{\left[G_{\rm RL}^{+}\left(\tau,\tau\right)t-G_{\rm
LR}^{+}(\tau,\tau)t^{\dagger}\right]{\bm\sigma}\right\}.
\end{equation}
Introducing the Fourier transform $G^{+}(\omega)$ of
$G^{+}(\tau,\tau^{\prime})$ , which is a function of
$\tau-\tau^{\prime}$, we have
\begin{equation}\label{uffa2}
\langle{\bf j}_{n-1}\rangle=\frac12\sum_{{\bf k}_{\parallel}}
\int\frac{{\rm d}\omega}{2\pi}{\rm Tr}\left\{\left[G_{\rm
RL}^{+}\left(\omega\right)t-G_{\rm
LR}^{+}(\omega)t^{\dagger}\right]{\bm\sigma}\right\}.
\end{equation}
The charge current is given by Eq. (\ref{uffa2}) with
$\frac12{\bm\sigma}$ replaced by the unit matrix multiplied by
$e/\hbar$.

Following Keldysh \cite{11,12} we now write
\begin{equation}\label{uffa3}
G_{\rm AB}^{+}(\omega)=\frac12\left(F_{\rm AB}+G_{\rm AB}^{\rm
a}-G_{\rm AB}^{\rm r}\right),
\end{equation}
where the suffices $A$ and $B$ are either $R$ or $L$. $F_{\rm
AB}(\omega)$ is the Fourier transform of 
\begin{equation}\label{uffa4}
F_{\rm AB}(\tau,\tau^{\prime})=-{\rm i}\langle[c_{\rm A}(\tau),c_{\rm
B}^{\dagger}(\tau^{\prime})]_{-}\rangle
\end{equation}
and $G^{\rm a}$, $G^{\rm r}$ are the usual advanced and retarded
Green functions \cite{54}. Note that in \cite{11} and \cite{12} the definitions
of $G^{\rm a}$ and $G^{\rm r}$ are interchanged and that in the
Green function matrix defined by these authors $G^{+}$ and $G^{-}$
should be interchanged.

Charge and spin current are related by Eqs. (\ref{uffa2}) and (\ref{uffa3}) to
the quantities $G^{\rm a}$, $G^{\rm r}$ and $F_{\rm AB}$. The latter
are calculated for the coupled system by starting with decoupled
left and right systems, each in equilibrium, and turning on the
hopping between planes L and R as a perturbation. Hence, we express
$G^{\rm a}$, $G^{\rm r}$ and $F_{\rm AB}$ in terms of retarded
surface Green functions $g_{L}\equiv g_{\rm LL}$, $g_{\rm R}\equiv
g_{\rm RR}$ for the decoupled equilibrium system. It is then found
\cite{13} that the spin current between the planes $n-1$ and $n$ can
be written as the sum $\langle{\bf j}_{n-1}\rangle=\langle{\bf
j}_{n-1}\rangle_1+\langle{\bf j}_{n-1}\rangle_2$, where the two
contributions to the spin current $\langle{\bf j}_{n}\rangle_1$,
$\langle{\bf j}_{n}\rangle_2$ are given by 
\begin{equation}\label{nonloso1}
\langle{\rm j}_{n-1}\rangle_1=\frac1{4\pi}\sum_{{\bf
k}_\parallel}\int{\rm d}\omega\Re{\rm Tr}[(B-A){\bm\sigma}]
[f(\omega-\mu_{\rm L})+f(\omega-\mu_{\rm R})].
\end{equation}

\begin{equation}\label{nonloso2}
\langle{\rm j}_{n-1}\rangle_2=\frac1{2\pi}\sum_{{\bf
k}_\parallel}\int{\rm d}\omega\Re{\rm Tr}\left\{[g_{\rm L}tABg_{\rm
R}^{\dagger}t^{\dagger}-AB+\frac12(A+B)]{\bm\sigma}\right\}
[f(\omega-\mu_{\rm L})-f(\omega-\mu_{\rm R})].
\end{equation}
Here, $A=[1-g_{\rm R}t^{\dagger}g_{\rm L}t]^{-1}$,
$B=[1-g^{\dagger}_{\rm R}t^{\dagger}g^{\dagger}_{\rm L}t]^{-1}$, and
 as in Section \ref{due} $f(\omega-\mu)$ is the Fermi function with
 chemical potential $\mu$ and $\mu_{\rm L}-\mu_{\rm R}=eV_{\rm b}$.
 In the linear-response case of small bias which we are considering,
 the Fermi functions in Eq. (\ref{nonloso2}) are expanded to first
 order in $V_{\rm b}$. Hence the energy integral is avoided, being
 equivalent to multiplying the integrand by $eV_{\rm b}$ and
 evaluating it at  the common zero-bias chemical potential $\mu_0$.

It can be seen that Eqs. (\ref{nonloso1}) and (\ref{nonloso2}), which determine
the spin and the charge currents, depend on just two quantities,
{\it i.e.} the surface retarded one-electron Green functions for a
system cleaved between two neighbouring atomic planes. The surface
Green functions can be determined without any approximations by the
standard adlayer method (see {\it e.g.} \cite{42,44}) for a fully
realistic band structure.

We first note that there is a close correspondence between Eqs.
(\ref{nonloso1}), (\ref{nonloso2}) and the generalised Landauer formula
(\ref{landau33}). The first term in Eq. (\ref{landau33}) corresponds to the
zero-bias spin current $\langle{\bf j}_{n-1}\rangle_1$ given by
Eq. (\ref{nonloso1}). When the cleavage plane is taken in the
spacer, the spin current $\langle{\bf j}_{n-1}\rangle_1$ determines
the oscillatory exchange coupling between the two magnets and it is
easy to verify that the formula for the exchange coupling obtained
from Eq. (\ref{nonloso1}) is equivalent to the formula used in
previous total energy calculations of this effect \cite{42,44}. The
contribution to the transport spin current given by Eq.
(\ref{nonloso2}) clearly corresponds to the second term in the
Landauer formula (\ref{landau33}) which is proportional to the bias
in the linear response limit. Placing the cleavage plane first
between any two neighbouring atomic planes in the spacer and then
between any two neighbouring planes in the lead, we obtain from Eq.
(\ref{nonloso2}) the total spin-transfer torque ${\bf T}^{\rm s-t}$
of Eq. (\ref{torque}) in Section \ref{due}.

The equivalence of the Keldysh and Landauer methods has been
demonstrated by calculating the currents
(\ref{nonloso1}) and (\ref{nonloso2}) analytically for the simple single
orbital model of Section \ref{due}. The results of that section,
such as Eq. (\ref{torquex})  are reproduced \cite{13}.

\section{Quantitative results for 
C\lowercase{o}/C\lowercase{u}/C\lowercase{o}(111)}\label{cinque}

We now discuss the application of the Keldysh formalism to a real
system. In particular we consider a realistic multiorbital model of
fcc Co/Cu/Co(111) with tight-binding parameters fitted to the
results of the first-principles band structure calculations, as
described previously \cite{42,44}.

\begin{figure}
\includegraphics[width=12cm]{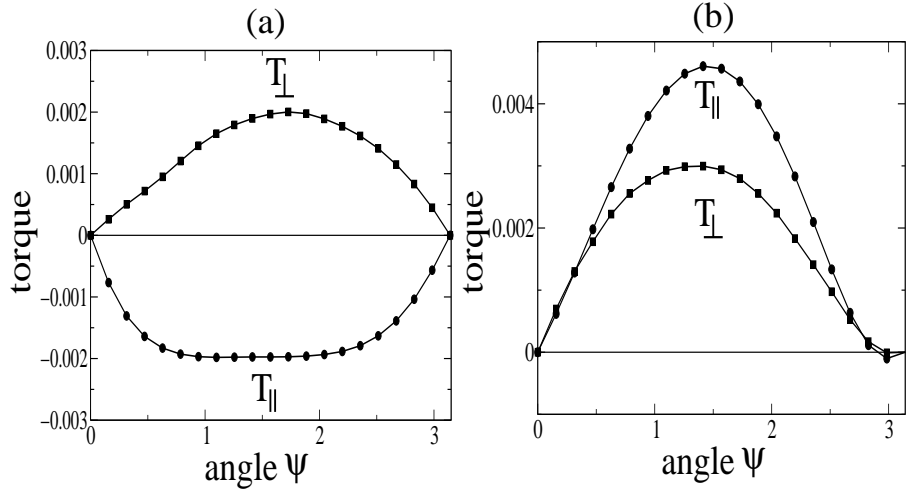}
\caption{Dependence of the spin-transfer torque $T_{\parallel}$ and
$T_{\perp}$ for Co/Cu/Co(111) on the angle $\psi$. The torques per
surface atom are in units of $eV_{\rm b}$. Figure (a) is for $M=1$,
and (b) for $M=2$ monolayers of Co in the switching magnet. }
 \label{fig13}
\end{figure}

Referring to Fig. \ref{fig1}, the system considered by Edwards {\it
et al.} \cite{13} consists of a semi-infinite slab of Co (polarising
magnet), the spacer of 20 atomic planes of Cu, the switching magnet
containing $M$ atomic planes of Co, and the lead which is
semi-infinite Cu. The spacer thickness of 20 atomic planes of Cu was
chosen so that the contribution of the oscillatory exchange coupling
term is so small that it can be neglected. The spin currents in the
right lead and in the spacer were determined from Eq.
(\ref{nonloso2}).
Figure \ref{fig13}(a),(b) shows the angular dependences of
$T_{\parallel}$, $T_{\perp}$ for the cases $M=1$ and $M=2$.
respectively.
For the monolayer switching magnet, the torques $T_{\parallel}$ and
$T_{\perp}$ are equal in magnitude and they have opposite sign.
However, for $M=2$, the torques have the same sign and $T_{\perp}$
is somewhat smaller than $T_{\parallel}$. A negative sign of the
ratio of the two torque components has important and unexpected
consequences for hysteresis loops as already discussed in Section
\ref{tre}. It can be seen that the angular dependence of both torque
components is dominated by a $\sin\psi$ factor but distortions from
this dependence are clearly visible. In particular, the slopes at
$\psi=0$ and $\psi=\pi$ are quite different. As pointed out in
Section \ref{tre}, this is important in the discussion of the
stability of steady states and leads to quite different magnitudes
of the critical biases $V_{\rm P}\rightarrow V_{\rm AP}$ and $V_{\rm
AP}\rightarrow V_{\rm P}$.

\begin{figure}
\includegraphics[width=8cm]{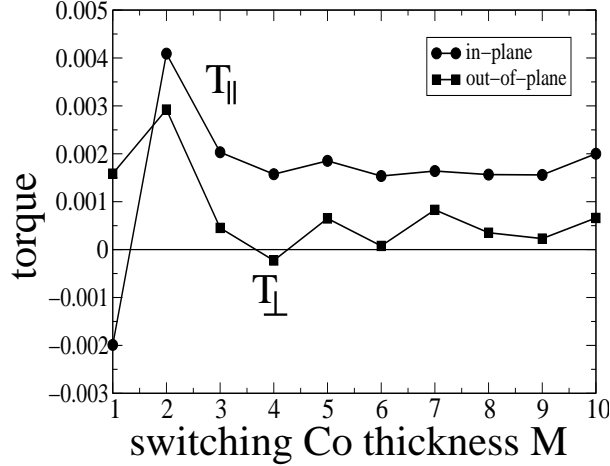}
\caption{Dependence of the spin-transfer torque $T_{\parallel}$ and
$T_{\perp}$ for Co/Cu/Co(111) on the thickness of the switching
magnet $M$ for $\psi=\pi/3$. The torques are in units of $eV_{\rm
b}$.}
 \label{fig14}
\end{figure}

In Fig. \ref{fig14} we reproduce the dependence of $T_{\perp}$ and
$T_{\parallel}$ on the thickness of the Co switching magnet. It can
be seen that the out-of-plane torque $T_{\perp}$ becomes smaller
than $T_{\parallel}$ for thicker switching magnets.
However, $T_{\perp}$ is by no means negligible (27$\%$ of
$T_{\parallel}$) even for a typical experimental thickness of the
switching Co layer of ten atomic planes. It is also interesting that
beyond the monolayer thickness, the ratio of the two torques is
positive with the exception of $M=4$.

The microscopically calculated spin-transfer torques for
Co/Cu/Co(111) were used by Edwards {\it et al.} \cite{13} as an
input into the phenomenological LLG equation. For simplicity the
torques as functions of $\psi$ were approximated by sine curves but
this is not essential. The LLG equation was first solved numerically
to determine all the steady states and then the stability discussion
outlined in the phenomenological section was applied to determine
the critical bias for which instabilities occur. Finally, the
ballistic resistance of the structure was evaluated from the
real-space Kubo formula at every point of the steady state path.
Such a calculation for the realistic Co/Cu system then gives
hysteresis loops of the resistance versus bias which can be compared
with the observed hysteresis loops. The LLG equation was solved
including a strong easy-plane anisotropy with $h_{\rm p}=100$. If we
take $H_{\rm u0}=1.86\times 10^9$sec$^{-1}$, corresponding to a
uniaxial anisotropy field of about 0.01T, this value of $h_{\rm p}$
corresponds to the shape anisotropy for a magnetisation of
$1.6\times 10^6$A/m, similar to that of Co \cite{sun}. Also a
realistic value \cite{sun} of the Gilbert damping parameter
$\gamma=0.01$ was used. Finally, referring to the geometry of Fig.
\ref{fig1}, two different values of the angle $\theta$ were employed
in these calculations: $\theta=2$rad and $\theta=3$rad, the latter
value being close to the value of $\pi$ which is realised in most
experiments.

\begin{figure}
\includegraphics[width=13cm]{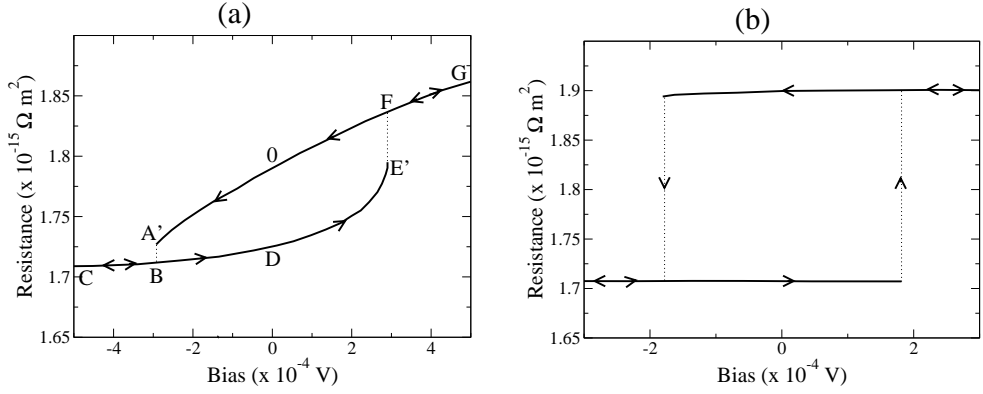}
\caption{Resistance of the Co/Cu/Co(111) junction as a function of
the applied bias, with $M=2$ monolayers of Co in the switching
magnet. (a) is for $\theta=2$rad and (b) for $\theta=3$rad.}
 \label{fig16}
\end{figure}

We first reproduce in Fig. \ref{fig16} the hysteresis loops for the
case of Co switching magnet consisting of two atomic planes. We note
that the ratio $r=T_{\perp}/T_{\parallel}\approx 0.65$ deduced from
Fig. \ref{fig13} is positive in this case. Fig. \ref{fig16}(a) shows
the hysteresis loop for $\theta=2$ and Fig. \ref{fig16}(b) that
for $\theta=3$.
The hysteresis loop for $\theta=3$ shown in Fig. \ref{fig16}(b)
 is an illustration of the stability scenario in zero applied field
 with $r>0$ discussed in Section \ref{tre}. As pointed out there the
 hysteresis curve is that of Fig. \ref{fig5}(a) which agrees with
 Fig. \ref{fig16}(b)
 when we remember that the reduced bias used in Fig. \ref{fig5} has
 the opposite sign from the bias in volts used in Fig. \ref{fig16}.
 It is rather interesting that the critical bias for switching is $\approx
 0.2$mV both for $\theta=2$ and $\theta=3$. When this bias
 is converted to the current density using the calculated ballistic
 resistance of the junction, it is found \cite{13} that the critical
 current for switching is $\approx 10^7$A/cm$^2$, which is in very
 good agreement with experiments \cite{albert}.

\begin{figure}
\includegraphics[width=13cm]{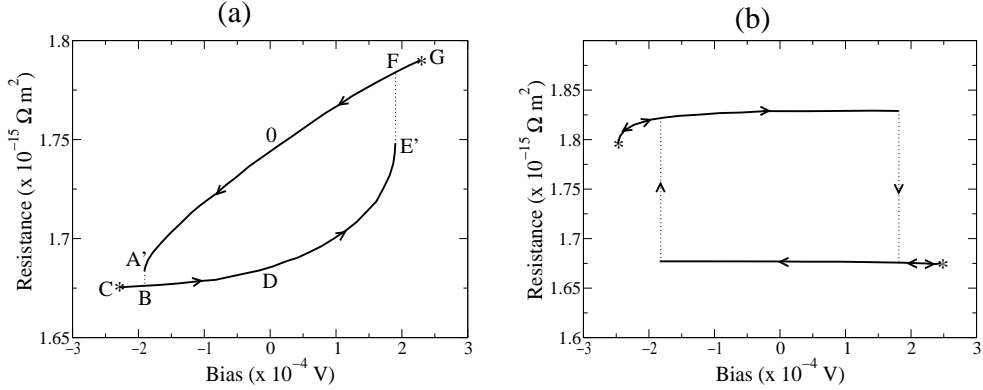}
\caption{Resistance of the Co/Cu/Co(111) junction as a function of
the applied current, with $M=1$ monolayer of Co in the switching
magnet. (a) is for $\theta=2$rad and (b) for $\theta=3$rad.}
 \label{fig17}
\end{figure}

 The hysteresis loops for the case of the Co switching magnet
 consisting of a single atomic plane are reproduced in Fig.
 \ref{fig17}. The values of $h_{\rm p}$, $\gamma$, $H_{\rm u0}$, and
 $\theta$ are the same as in the previous example.
However the ratio $r\approx -1$ is now negative and the hysteresis
loops in Fig. \ref{fig17} illustrate the interesting behaviour
discussed in Section \ref{tre} when the system subjected to a bias
higher than a critical bias moves to the ''both unstable" region
shown in Fig. \ref{fig6}. 
As in Fig. \ref{fig7} the points on the hysteresis loop in
Fig. \ref{fig17} corresponding to the critical bias are labelled by asterisks.
Fig. \ref{fig17}(b)
and Fig. \ref{fig7}(a) are in close correspondence because Fig.
\ref{fig7}(a) is for $r_{\rm c}<r<0$ and in the present case $r=-1$,
$r_{\rm c}=-2/(\gamma h_{\rm p})=-2$. Also, from Fig.
\ref{fig13}(a), $g_{\parallel}<0$ so that $vg_{\parallel}$ in Fig.
\ref{fig7}(a) has the same sign as the voltage $V$ in Fig.
\ref{fig17}(b).

\section{Summary}

Spin-transfer torque is responsible for current-driven switching of
magnetisation in magnetic layered structures. The simplest
theoretical scheme for calculating spin-transfer torque is a
generalised Landauer method and this is used in Section \ref{due} to
obtain analytical results for a simple model. The general
phenomenological form of spin-transfer torque is deduced in Section
\ref{tre} and this is introduced into the Landau-Lifshitz-Gilbert
equation, together with torques due to anisotropy fields. This
describes the motion of the magnetisation of the switching magnet
and the stability of the steady states (constant current and
stationary magnetisation direction) is studied under different
experimental conditions, with and without external field. This leads
to hysteretic and reversible behaviour in resistance versus bias (or
current) plots in agreement with a wide range of
experimental observations. In Section \ref{quattro} the general
principles of a self-consistent treatment of spin-transfer torque are
discussed and the Keldysh formalism for quantitative calculations is
introduced. This approach to the non-equilibrium problem of electron
transport uses Green functions which are very convenient to
calculate for a realistic multiorbital tight-binding model of the
layered-structure. In Section \ref{cinque} quantitative calculations
for Co/Cu/Co(111) systems are presented which yield switching
currents of the observed magnitude.

This study of current-driven switching of magnetisation was carried
out in collaboration with J. Mathon and A. Umerski and financial
support was provided by the UK Engineering and Physical Research
Council (EPSRC).


\begin{thebibliography}{99}

\bibitem{1} P. Gr\"{u}nberg, R. Schreiber, Y. Pang, M. B. Brodsky, and
  H. Sower, Phys. Rev. Lett. {\textbf 57}, 2442 (1986); M. N. Baibich,
  J. M. Broto, A. Fert, Van Dau Nguyen, F. Petroff, P. Etienne,
  G. Creuset, A. Friederich, and J. Chazelas, Phys. Rev. Lett. {\bf
  61}, 2472 (1998)
\bibitem{2}S. S. P. Parkin {\it et al.}, J. Appl. Phys. {\bf 85},
  5828 (1999).
\bibitem{3}J. C. Slonczewski, J. Magn. Magn. Mater. {\bf 159}, L1 (1996).
\bibitem{4}X. Waintal, E. B. Myers, P. W. Brouwer, and D. C. Ralph,
  Phys. Rev. B {\bf 62}, 12317 (2000)
\bibitem{5}{\it e.g.} ``Transport in Nanostructures'' by D. K. Ferry
  and S. M. Goodnick (Cambridge University Press 1997).
\bibitem{sun} J. Z. Sun, Phys. Rev. B {\bf 62}, 570 (2000).
\bibitem{albert} F. J. Albert, J. A. Katine, R. A. Buhrman, and
  D. C. Ralph, Appl. Phys. Lett. {\bf 77}, 3809 (2000).
\bibitem{16} J. Grollier, V. Cross, A. Hamzic, J. M. George,
  H. Jaffres, A. Fert, G. Faini, J. Ben Youssef, and H. Le Gall,
  Appl. Phys. Lett. {\bf 78}, 3663 (2001).
\bibitem{17}E. C. Stoner and E. P. Wohlfarth, Phil, Trans. Roy. Soc. A
  {\bf 240}, 599 (1948).
\bibitem{18}J. Grollier, V. Cross, H. Jaffres, A. Hamzic, J. M. George,
  G. Faini, J. Ben Youssef, H. Le Gall, and A. Fert,
   Phys. Rev. B {\bf 67}, 174402 (2003).
\bibitem{19} F. J. Albert, N. C. Emley, E. B. Myers, D. C. Ralph, and
  R. A. Buhrman, Phys. Rev. Lett. {\bf 89}, 226802 (2002).
\bibitem{20} D. W. Jordan and P. Smith, Nonlinear Ordinary
  Differential Equations, Clarendon Press, Oxford (1977).
\bibitem{13} D. M. Edwards, F. Federici, G. Mathon, and A. Umerski,
  Phys. Rev. B {\bf 71}, 134501 (2005).
\bibitem{21} J.A. Katine, F. J. Albert, R. A. Buhrman, E. B. Myers, and
  D. C. Ralph,  Phys. Rev. Lett. {\bf 84}, 3149 (2000).
\bibitem{22} S. I. Kiselev, J. C. Sankey, I. N. Krivorotov,
  N. C. Emley, R. J. Schoelkopf, R. A. Buhrman, and D. C. Ralph, Nature {\bf 425}, 380 (2003).
\bibitem{23} S. Urazhdin, N. O. Birge, W. P. Pratt, Jr., and J. Bass, Phys. Rev. Lett. {\bf 91}, 146803 (2003)
\bibitem{24} M. R. Pufall, W. H. Rippard, S. Kaka, S. E. Russek,
  T. J. Silva, J. Katine, and M. Carey,  Phy. Rev. Lett. {\bf 69},
  214409 (2004).
\bibitem{25} E. B. Myers, F. J. Albert, J. C. Sankey, E. Bonet,
  R. A. Buhrman, and D. C. Ralph, Phys. Rev. Lett. {\bf 89}, 196801 (2002)
\bibitem{26} M. A. Zimmler, B. \"{O}zyilmaz, W. Chen, A. D. Kent,
  J. Z. Sun, M. J. Rooks, and R. H. Koch,  Phy. Rev. B {\bf 70},
  184438 (2004).
\bibitem{27}M. Tsoi, A. G. M. Jansen, J. Bass, W. C. Chiang. M. Seck, V.Tsoi,
  and P. Wyder, Phys. Rev. Lett. {\bf 80}, 4281 (1998).
\bibitem{28}M. Tsoi, A. G. M. Jansen, J. Bass, W. C. Chiang. V. Tsoi,
  and P. Wyder, Nature {\bf 406}, 46 (2000)
\bibitem{29} E. B. Myers, D. C. Ralph, J. A. Katine, R. N. Louie and
  R. A. Buhrman, Science {\bf 285}, 867 (1999).
 \bibitem{6} J. C. Slonczewski, J. Magn. Magn. Mater. {\bf 195}, L261 (1999);
  {\bf 247}, 324 (2002).
\bibitem{30}B. \"{O}zyilmaz, A. D. Kent, D. Monsma, J. Z. Sun,
  M. J. Rooks, and R. H. Koch, Phys. Rev. Lett. {\bf 91}, 067203 (2003).
\bibitem{31} M. Tsoi, J. Z. Sun,
  M. J. Rooks, R. H. Koch, and S. S. P. Parkin, Phys. Rev. B {\bf 69}, 100406(R) (2004).
\bibitem{32}W. F. Brown, Phys. Rev. B {\bf 130}, 1677 (1963).
\bibitem{33}Z. Li and S. Zhang, Phys. Rev. B {\bf 69}, 134416 (2004).
\bibitem{34}J. Miltat, G. Albuquerque, A. Thiaville, and C. Vouille, J. Appl.
Phys. {\bf 89}, 6982 (2001).
\bibitem{35}Z. Li and S. Zhang, Phys. Rev. B {\bf 68}, 024404 (2003).
\bibitem{36} C. R. P. Cowburn, C. K. Koltsov, A. O. Adeyeye, and M.
E. Welland, Phys. Rev. Lett. {\bf 83}, 1042 (1999).
\bibitem{37} Y. Jiang, S. Abe, T. Ochiai, T. Nozaki, A. Hirohata, N.
Tezuka, and K. Inomata, Phys. Rev. Lett. {\bf 92}, 167204 (2004).
\bibitem{38} Y. Jiang, T. Nozaki, S. Abe, T. Ochiai, A. Hirohata, N.
Tezuka, and K. Inomata, Nature Materials {\bf 3}, 361 (2004).
\bibitem{39} N. Tezuka (private communication)
\bibitem{40} K. Capelle and B. L. Gyorffy, Europhys. Lett. {\bf 61},
354 (2003).
\bibitem{41} J. Sun, Nature {\bf 424}, 359 (2003).
\bibitem{42} J. Mathon, Murielle Villeret, A. Umerski, R. B. Muniz,
J. d'Albuquerque e Castro, and D. M. Edwards, Phys. Rev B, {\bf 56},
11797 (1997).
\bibitem{44} J. Mathon, Murielle Villeret, R. B. Muniz, J. 
d'Albuquerque e Castro, and D. M. Edwards, Phys. Rev. Lett. {\bf
74}, 3696 (1995).
\bibitem{46}S. Zhang, P. M. Levy, and A. Fert , Phys. Rev. Lett. {\bf
88},236601 (2002).
\bibitem{47} A. A. Kovalev, A. Brataas, and G. E. W. Bauer, Phys.
Rev. B {\bf 66 }, 224424 (2002).
\bibitem{em} D. M. Edwards and J. Mathon in ''Nanomagnetism: Multilayers,
Ultrathin Films and Textured Media", eds. J. A. C. Bland and D. L.
Mills (Elsevier, to be published).
\bibitem{11} L. V. Keldysh, Sov. Phys. JETP, {\bf 20}, 1018 (1965).
\bibitem{12} C. Caroli, R. Combescot, P. Nozieres, and D.
Saint-James, J. Phys. C {\bf 4}, 916 (1971).
\bibitem{53}D. M. Edwards in: Exotic states in quantum
nanostructures, ed. by S. Sarkar, Kluwer Academic Press (2002).
\bibitem{54} G. D. Mahan, Many Particle Physics, 2nd Ed., Plenum
Press, New York (1990).
\end{thebibliography}
\end{document}